\documentclass[a4paper, 11pt, oneside]{article}
\pdfoutput=1
\usepackage{jheppub}

\usepackage{graphicx,wrapfig,float,slashed,cancel}
\usepackage{amsmath,amssymb,epsfig,graphicx,xcolor}
\usepackage{mathrsfs}
\usepackage{multirow}
\usepackage{epstopdf}
\usepackage{soul}
\usepackage{ragged2e}
\usepackage{lipsum}
\usepackage{tabu}
\allowdisplaybreaks
\usepackage{pifont}
\usepackage[normalem]{ulem}
\usepackage{bbm}
\usepackage{subcaption}
\usepackage[force]{feynmp-auto}
\usepackage{booktabs}
\newcommand{\nc}{\newcommand}
\nc{\non}{\nonumber}
\nc{\hc}{\hbox {h.c.}}
\nc{\noi}{\noindent}
\nc{\barx}{\bar{x}}
\nc{\pbarn}{\;\hbox {pb}}
\nc{\fbarn}{\;\hbox {fb}}

\nc{\hsp}{\hspace{0.4cm}}
\nc{\lsp}{\hspace{0.8cm}}
\nc{\Lsp}{\hspace{1.6cm}}
\nc{\LLsp}{\lsp\lsp}
\nc{\lra}{\longrightarrow}
\nc{\p}{\prime}
\nc{\sgn}{\text{sgn}}
\nc{\ph}{\varphi}
\nc{\op}{{\cal O}}
\nc{\eq}{\text{Eq.~}}
\nc{\cL}{\mathcal{L}}
\nc{\mo}{\texttt{micrOMEGAs }}
\nc{\vmol}{v_{\text{M\o l}}}
\nc{\eff}{\text{eff}}
\nc{\sm}{\text{\rm SM}}
\nc{\mchi}{m_\chi}
\nc{\mtchi}{m_{\tilde\chi}}
\nc{\mphi}{m_\phi}
\nc{\mtphi}{m_{\tilde\phi}}
\nc{\beq}{\begin{equation}}  \nc{\eeq}{\end{equation}}
\nc{\bea}{\begin{eqnarray}}  \nc{\eea}{\end{eqnarray}}
\nc{\baa}{\begin{array}}     \nc{\eaa}{\end{array}}
\nc{\bit}{\begin{itemize}}   \nc{\eit}{\end{itemize}}
\nc{\ben}{\begin{enumerate}} \nc{\een}{\end{enumerate}}
\nc{\bce}{\begin{center}}    \nc{\ece}{\end{center}}
\nc{\bpm}{\begin{pmatrix}}   \nc{\epm}{\end{pmatrix}}
\nc{\bvt}{\begin{verbatim}}  \nc{\evt}{\end{verbatim}}
\nc{\sdm}{\text{\tiny SDM}}
\nc{\vdm}{\text{\tiny VDM}}

\nc{\te}{}

\def\lsim{\mathrel{\raise.3ex\hbox{$<$\kern-.75em\lower1ex\hbox{$\sim$}}}}
\def\gsim{\mathrel{\raise.3ex\hbox{$>$\kern-.75em\lower1ex\hbox{$\sim$}}}}

\renewcommand{\Re}{\mbox{Re\thinspace}}
\renewcommand{\Im}{\mbox{Im\thinspace}}
\def\udots{\mathinner{\mkern1mu\raise1pt\vbox{\kern7pt\hbox{.}}\mkern2mu\raise4pt\hbox{.}\mkern2mu\raise7pt\hbox{.}\mkern1mu}}

\def\gev{\;\text{GeV}}
\def\tev{\;\text{TeV}}

\def\zBB{{\mathbbm Z}}
\def\z2{\zBB_2}
\def\lamh{\lambda_H}
\def\lams{\lambda_S}
\def\lamsm{\lambda_{SM}}
\def\z2{\zBB_2}

\def\mone{m_1}
\def\mtwo{m_2}
\def\vs{v_S}
\def\va{v_A}
\def\gx{g_X}

\def\uone{U(1)_X}
\def\mdm{m_{DM}}
\def\inv#1{\frac1{#1}}
\title{Testing scalar versus vector dark matter}
\author[2]{Duarte Azevedo,}
\author[1]{Mateusz Duch,}
\author[1]{Bohdan Grzadkowski,} 
\author[1]{Da Huang,} 
\author[1]{Michal Iglicki,} 
\author[2]{and Rui Santos}
\affiliation[1]{Faculty of Physics, University of Warsaw,
Pasteura 5, 02-093 Warsaw, Poland}
\affiliation[2]{
Centro de F\'{\i}sica Te\'{o}rica e Computacional, Faculdade de Ci\^{e}ncias, Universidade de Lisboa,
Campo Grande, Edif\'{\i}cio C8 1749-016 Lisboa, Portugal}
\emailAdd{dazevedo@alunos.fc.ul.pt}
\emailAdd{mateusz.duch@fuw.edu.pl}
\emailAdd{bohdan.grzadkowski@fuw.edu.pl}
\emailAdd{da.huang@fuw.edu.pl}
\emailAdd{Michal.Iglicki@fuw.edu.pl}
\emailAdd{rasantos@fc.ul.pt}

\abstract{
We investigate and compare two simple models of dark matter (DM): a vector and a scalar DM model. Both models require the presence of two
physical Higgs bosons $h_1$ and $h_2$ which come from mixed components of the standard Higgs doublet $H$ and a complex singlet $S$. In the Vector model, the extra $U(1)$ symmetry is spontaneously broken by the vacuum of the complex field $S$. This leads to a massive gauge boson $X^\mu$ that is a DM candidate stabilized by the dark charge conjugation symmetry $S \to S^*$, $X^\mu\to -X^\mu$.
On the other hand, in the Scalar model the gauge group remains the standard one.
The DM field $A$ is the imaginary component of $S$ and the stabilizing symmetry is also the dark charge conjugation $S \to S^*$ ($A \to - A$).
In this case, in order to avoid spontaneous breaking, the $U(1)$ symmetry is broken explicitly, but softly, 
in the scalar potential. The possibility to disentangle the two models has been investigated. We have analyzed collider, cosmological, DM direct and indirect detection constraints and shown that there are regions in the space spanned by the mass of the non-standard Higgs boson and the mass of the DM particle where the experimental bounds exclude one of the models. We have also considered possibility to disentangle the models at $e^+e^-$ collider and concluded that the process $e^+e^-\to Z + \text{DM}$ provides a useful tool to distinguish the models.
}


\begin{document}

\maketitle
\flushbottom

\section{Introduction} 
\label{intro}

The Higgs boson was discovered at CERN's Large Hadron Collider (LHC) by the ATLAS~\cite{ATLAS:2012ae} and CMS~\cite{Chatrchyan:2012tx} collaborations
thus turning one important page in our knowledge of the Universe by not only discovering a new particle but also to hint very strongly that there is a 
mechanism of spontaneous symmetry breaking giving mass to both gauge bosons and fermions. Over the years, it has become increasingly clear
that this boson resembles very much the one predicted by the Standard Model (SM). However, there are still many unsolved problems in particle physics that are not answered by the SM. One of them is the existence of the dark matter (DM) in the universe which presence cannot be attributed to any known particles.

Although the measurements of the Higgs couplings are  quite demanding for the so-called Beyond the Standard Model (BSM) models,
there is still plenty of space in the present results to include new physics. Some of the BSM models can be compatible with the measurements while providing 
solutions to some of the outstanding questions of particle physics. Such is the case of the models studied in this work. Both the extension with a complex 
singlet~\cite{Silveira:1985rk, McDonald:1993ex, Burgess:2000yq, Bento:2000ah, Gonderinger:2012rd, Gabrielli:2013hma}  
and the extension with a new Abelian vector boson together with a complex singlet~\cite{Hambye:2008bq,Lebedev:2011iq,Farzan:2012hh,Baek:2012se,Baek:2014jga,Duch:2015jta}  provide DM candidates still compatible with collider bounds, and direct or
indirect detection experiments. The models can also undergo a strong first-order phase transition during the era of 
EWSB~\cite{Profumo:2014opa, Barger:2011vm, Espinosa:2011ax, McDonald:1993ey, Branco:1998yk, Barger:2008jx, Gonderinger:2012rd, Jiang:2015cwa, Chao:2017oux, Chao:2014ina} thus explaining electroweak baryogenesis

Extra scalar singlets are dimension one fields and therefore prone to couple to the SM scalar sector in a renormalisable way without any suppression by inverse powers of the scale of BSM,  a concept introduced 
in~\cite{Patt:2006fw} and known as the {\em Higgs portal}. Assuming the scale of new physics is the GUT or the Planck scale
we are at present bound to work with minimal theories that are valid up to high energy scales. This theory has in particular
to be stable under  the Renormalization Group Evolution (RGE) which is an issue already in the SM. The measurement
of the Higgs and the top-quark masses show that the SM is either in a marginally stable or in a metastable region of the parameter space~\cite{Degrassi:2012ry, Alekhin:2012py}.
However, as shown at two-loop level in~\cite{Costa:2014qga,Duch:2015jta}, these models not only provide a DM candidate but they also improve the stability of the SM and present a posibility to solve the baryon asymmetry problem. 

In this article we explore possibilities of distinguishing the scalar and the vector DM (VDM) models. The minimal VDM requires an extra $U(1)$ gauge symmetry that is spontaneously broken by a vacuum expectation value (vev) of a complex scalar neutral field under the SM symmetries but charged under the extra $U(1)$. This model bears many similarities with a model of scalar DM (SDM) which is a component of an extra complex scalar field (that develops a vev) which is added to the SM. In both cases there are two scalar physical Higgs bosons $h_{1,2}$ that mix in the scalar mass matrix with a mixing angle $\alpha$. So the goal of this paper is to investigate if those two models could be distinguished. This is a very pragmatic task, both models are attractive candidates for simple DM theories, therefore it is worth knowing if there are observables which can distinguish them. 

Using the~\textsc{ScannerS} program~\cite{Coimbra:2013qq} we impose
the most relevant bounds: theoretical, collider experiment bounds, precision electroweak physics, DM direct and indirect detection experiments and DM relic density.
The parameter space of each model is scanned with all the above constraints providing the regions of the parameter space where the models can indeed be distinguished. Whenever possible
these results are presented in terms of physical observables that can be measured at the LHC. Finally we present a direct way to distinguishing the models by looking at the 
energy distribution in Higgs associated production, with the Higgs decaying to DM, at a future electron-positron collider. 

The paper is organized as follows. In Sec.~\ref{SDM}  we present the complex singlet extension of the SM, reviewing its main properties and setting notation. In Sec.~\ref{DD_dd_sdm} and \ref{inv_dec_sdm} we discuss the scattering of scalar DM off nuclei and invisible SM-like Higgs boson decays, respectively.
In Sec.~\ref{VDM} we set the review of most relevant aspects of the vector DM model. In Sec.~\ref{DD_dd_vdm}  and Sec.~\ref{inv_dec_vdm} constraints from DM direct detection and invisible decays of SM-like Higgs boson are formulated, respectively.
In Sect.~\ref{coll_tests} we present a discussion of the possibility to distinguish the models at a future electron-positron collider. The results of the scan showing the allowed parameter
space for each model are presented in Sec.~\ref{results}. In the conclusions, Sec.~\ref{sec:conclusion}, we summarize our findings. Technical details concerning Goldstone Boson couplings to Higgs bosons are left to the appendices.

\section{Scalar Dark Matter}
\label{SDM}

Gauge singlet scalars as candidates for DM were first proposed in \cite{Silveira:1985rk} and \cite{McDonald:1993ex} and then discussed by many authors.
Even though the minimal model of scalar DM assumes merely an addition of a real scalar field odd under a $\z2$ symmetry,
here we are going to consider a model that requires an extension by a complex scalar filed $S$. The motivation is to compare the VDM with a SDM that are in some sense similar. In order to stabilize a component of $S$ we require an invariance under DM charge conjugation $C:\; S\rightarrow S^*$, which guarantees stability of the imaginary part of $S$, $A\equiv \Im S/\sqrt{2}$. The real part, $\phi_S \equiv \Re S/\sqrt{2}$, is going to develop a real vacuum expectation value $\langle \phi_S \rangle = \langle S \rangle = v_S/\sqrt{2}$.~\footnote{This is a choice that fixes the freedom (phase rotation of the complex scalar) of choosing a weak basis that could be adopted to formulate the model. The model is defined by symmetries imposed in this particular basis in which the scalar vacuum expectation value is real.} Therefore $\phi_S$ will mix with the neutral component of the SM Higgs doublet $H$, in exactly the same manner as it happens for the VDM. In order to simplify the potential we impose in addition a $\z2$ symmetry $S \rightarrow -S$, which eliminates odd powers of $S$.
Eventually the scalar potential reads:
\begin{equation}
	V=-\mu_H^2 |H|^2+\lambda_H |H|^4 -\mu_S^2|S|^2 + \lambda_S|S|^4+\kappa|S|^2|H|^2 + \mu^2 (S^2+S^{*\, 2})
	\label{pot_sdm}
\end{equation}
with $\mu^2$ real, as implied by the $C$ symmetry. 
Note that the $\mu^2$ term breaks the $U(1)$ explicitly, so the pseudo-Goldstone boson, $A$ is massive. In the limit of exact symmetry, $A$ would be just a genuine, massless Goldstone boson. Since the symmetry-breaking operator $\mu^2 (S^2+S^{*\, 2})$ is of dimension less that 4, its presence does not jeopardise renormalizability even if non-invariant higher dimension operators were not introduced, see for instance \cite{Pokorski:1987ed}. Note that dimension 3 terms are disallowed by the $\z2$'s and gauge symmetries. In other words, we can limit ourself to dimension 2 $U(1)$ breaking terms preserving the renormalizability of the model. The freedom to introduce solely the soft breaking operators offers a very efficient and economical way to generate mass for the pseudoscalar $A$ without the necessity to introduce dimension 4 terms like $S^4$ or $|S|^2S^2$ and keeping the renormalizability of the model.
It is also worth noticing that the $\z2$ symmetry $S \rightarrow -S$ is broken spontaneously by $v_S$ and therefore $\phi_S$, the real part of $S$, is not stable, making $A$ the only DM candidate.

The requirement of asymptotic positivity of the potential implies the following constraints that we impose in all further discussions:
\beq
\lambda_H > 0, \ \ \lambda_S >0, \ \ \kappa > -2 \sqrt{\lambda_H \lambda_S}.
\eeq
Hereafter the above conditions will be referred to as the positivity or stability conditions. 

The scalar fields can be expanded around the corresponding generic vev's as follows
\beq
S=\frac{1}{\sqrt{2}}(\vs + i \va + \phi_S +iA)  \ \ , \ \  H^0= \frac{1}{\sqrt{2}}(v + \phi_H+ i\sigma_H)   \ \  \text{where} \ \  H=\binom{H^+}{H^0},
\eeq
where we have temporarily allowed $\langle S \rangle$ to be complex.
 
Locations of extrema of the potential (\ref{pot_sdm}), corresponding values of the potential and corresponding curvatures in the basis $ \left(\phi_H, \phi_S, A \right)$ are as follows
\ben
\item[{\bf v1:}] 
\bea
v^2&=&\frac{4 \lambda_S \mu^2_H - 2\kappa (\mu^2_S-2 \mu^2)}{4\lambda_H\lambda_S-\kappa^2},\hsp
v_S^2=\frac{4 \lambda_H (\mu^2_S-2\mu^2) - 2\kappa \mu^2_H }{4\lambda_H\lambda_S-\kappa^2}, \hsp
v_A^2=0 \label{vac1}\\
V_1&=&\frac{-1}{4\lambda_H\lambda_S-\kappa^2}\left\{\lamh(\mu^2_S-2\mu^2)^2+\mu_H^2\left[\lams\mu_H^2-\kappa (\mu^2_S-2\mu^2)
\right]\right\} \label{V1}\\
\mathcal{M}^2 &=& \left( 
\begin{array}{ccc}
2 \lambda_H v^2  & \kappa v v_S & 0 \\ 
 \kappa v v_S   & 2 \lambda_S v^2_S & 0 \\
0 & 0 & - 4 \mu^2
\end{array} 
\right) \label{massv1},
\eea
\item[{\bf v2:}] 
\bea
v^2&=&\frac{4 \lambda_S \mu^2_H - 2\kappa (\mu^2_S+2 \mu^2)}{4\lambda_H\lambda_S-\kappa^2},\hsp
v_S^2=0,\hsp
v_A^2=\frac{4 \lambda_H (\mu^2_S+2\mu^2) - 2\kappa \mu^2_H }{4\lambda_H\lambda_S-\kappa^2}, 
\label{vac2}\\
V_2&=&\frac{-1}{4\lambda_H\lambda_S-\kappa^2}\left\{\lamh(\mu^2_S+2\mu^2)^2+\mu_H^2\left[\lams\mu_H^2-\kappa (\mu^2_S+2\mu^2)
\right]\right\} \label{V2} \\
\mathcal{M}^2 &=& \left( 
\begin{array}{ccc}
2 \lambda_H v^2  & 0 & \kappa v v_S \\ 
0 & 4 \mu^2 & 0  \\
\kappa v v_S & 0 & 2 \lambda_S v^2_S
\end{array} 
\right) \label{massv2},
\eea
\item[{\bf v3:}] 
\bea
v^2&=&\frac{\mu^2_H}{\lambda_H},\hsp v_S^2=0,\hsp v_A^2=0, \label{vac3} \\
V_3&=&-\frac{\mu_H^4}{4\lamh} \label{V3} \\
\mathcal{M}^2 &=& \left( 
\begin{array}{ccc}
2 \mu_H^2   & 0 & 0 \\ 
0 & 2 \mu^2+\frac{\kappa \mu_H^2}{2 \lamh} - \mu_S^2 & 0  \\
0 & 0 & -2 \mu^2+\frac{\kappa \mu_H^2}{2 \lamh} - \mu_S^2
\end{array} 
\right) \label{massv3},
\eea
\item[{\bf v4:}] 
\bea
v^2&=&0,\hsp v_S^2=\frac{\mu_S^2-2\mu^2}{\lambda_S},\hsp v_A^2=0, \label{vac4} \\
V_4&=&-\frac{(\mu^2_S-2\mu^2)^2}{4\lams} \label{V4}
\eea
\item[{\bf v5:}] 
\bea
v^2&=&0,\hsp v_S^2=0,\hsp v_A^2=\frac{\mu_S^2+2\mu^2}{\lambda_S}, \label{vac5} \\
V_5&=&-\frac{(\mu^2_S+2\mu^2)^2}{4\lams} \label{V5}
\eea
\een

Note that $v_S\neq 0$ and $v_A\neq 0$ may happen only if $\mu^2=0$. Since non-zero $\mu^2$ is essential to avoid the the appearance
of a Goldstone boson, we do not consider those points any more. 

Forcing the vacuum  {\bf v1}  to be the global minimum implies that we have to assume $\lamh>0$,  $4\lamh\lams-\kappa^2>0$ and $\mu^2<0$. 
Then for consistency we enforce the conditions
\beq
2\lams \mu_H^2 > \kappa (\mu_S^2-2\mu^2) \hsp \text{and} \hsp 2 \lamh (\mu_S^2-2\mu^2) > \kappa \mu_H^2
\label{con_con}
\eeq

It turns out that $V_1 < V_4$ for any choice of parameters, while $V_4< V_5$ for $\mu^2<0$.
From (\ref{con_con}) one can find that the vacuum {\bf v3} is never a minimum. Obviously, {\bf v2} is not a minimum either 
for $\mu^2<0$. Therefore we conclude that for $\mu^2<0$ the vacuum {\bf v1} is the global minimum. Note that in this case
$A$ is indeed a pseudo-Goldstone boson and its mass vanishes in the limit of exact global $U(1)$ as it was discussed and anticipated below (\ref{pot_sdm}).
The presence of the $U(1)$ breaking term $\mu^2(S^2+S^{*\,2})$ implies a trivial shift of the $\mu_S^2 \to \mu_S^2 - 2 \mu^2$ and an addition of the Goldstone boson mass $-4\mu^2$. In fact, an equivalent $U(1)$ breaking would be to add just the Goldstone boson mass without the trivial shift by replacing $\mu^2(S^2+S^{*\,2})$ by $\mu^2(S-S^*)^2$. 

Similar models have been considered in a more general context including a possibility of fast first order phase transition in \cite{Gonderinger:2012rd, Barger:2010yn, Barger:2008jx}. In the VDM that we consider here, $A$ becomes a longitudinal component of the massive DM vector $X$, but it remains an independent degree of freedom.

There are two mass eigenstates, $h_1$ and $h_2$, in this model. The mass matrix \ref{massv1} can be diagonalised by the orthogonal rotation matrix $R$ acting on the 
space spanned by the two CP-even scalars $\phi_H$ and $\phi_S$:
\begin{eqnarray}\label{mixing_s}
\left(\begin{array}{c}
\phi_H \\ \phi_S
\end{array}\right) = \left( \begin{array}{cc}
\cos\alpha & -\sin\alpha \\
\sin\alpha & \cos\alpha
\end{array} \right) \left( \begin{array}{c}
h_1 \\ h_2
\end{array}\right) \, .
\label{mix}
\end{eqnarray}
We assume hereafter that $h_1$ is the $125\gev$ boson observed at the LHC.

Note that the third spin-zero state $A$ does not mix with the former ones as the $\z2$ dark symmetry remains unbroken
by the real vev. We choose as independent parameters of the model the set: $v_S$, $\sin\alpha$, $m_2$ and $m_A$, while the parameters of the potential can be written as
functions of this independent set  and $v=246.22\gev$ and $m_1=125.09\gev$ as follows:
\begin{eqnarray}\label{relation_s}
\kappa = \frac{\sin 2\alpha (m_1^2 -m_2^2)}{2v v_S}, \,\, \lambda_S = \frac{\cos\alpha^2 m_2^2 + \sin\alpha^2 m_1^2}{2 v_S^2}, \,\, \lambda_H = \frac{\cos\alpha^2 m_1^2 + \sin\alpha^2 m_2^2}{2 v^2}\,.
\end{eqnarray}

The vertices relevant for the calculation of annihilation cross-section in the scalar DM model have been collected in tab.~\ref{SDM_ver}.
\begin{table}[h]
\begin{center}
\te
\begin{tabular}{|c|c|c|c|c|}
\hline
& &    & & \vspace{-0.5cm} 
\\
$ -i\frac{m_i^2}{\vs}R_{2i}$
& $i2m_X\gx R_{1i}$
& $i\frac{2M^2_{W}}{v}R_{1i}$
& $i\frac{M_{F}}{v}R_{1i}$
& $-i(R_{2i}R_{2j}\lambda_S-R_{1i}R_{1j}\kappa)$
\\
& &    & & \vspace{-0.5cm}
\\
\hline
& &    & & \vspace{-0.5cm} 
\\  
\begin{fmffile}{v1}
	        \begin{fmfgraph*}(60,50)
	            \fmfleft{i,j}
	            \fmfright{k}
	      \fmf{dashes,label=$A$,l.s=right}{i,v}
		    \fmf{dashes,label=$A$,l.s=left}{j,v}
		    \fmf{dashes,label=$h_i$}{v,k}
	        \end{fmfgraph*}
\end{fmffile} 
&
 \begin{fmffile}{v2}
	        \begin{fmfgraph*}(60,50)
	            \fmfleft{i,j}
	            \fmfright{k}
	            \fmf{boson,label=$Z$}{i,v}
		    \fmf{boson,label=$Z$}{j,v}
		    \fmf{dashes,label=$h_i$}{v,k}
	        \end{fmfgraph*}
\end{fmffile}  & 
 \begin{fmffile}{v3}
	        \begin{fmfgraph*}(60,50)
	            \fmfleft{i,j}
			\fmfright{k}
			\fmf{boson,label=$W^+$}{i,v}
			\fmf{boson,label=$W^-$}{j,v}
			\fmf{dashes,label=$h_i$}{v,k}
		  \end{fmfgraph*}
  \end{fmffile} &
  \begin{fmffile}{v4}
		  \begin{fmfgraph*}(60,50)
		      \fmfleft{i,j}
		      \fmfright{k}
		      \fmf{fermion,label=$\bar{f}$,l.s=left}{v,i}
		      \fmf{fermion,label=$f$,l.s=left}{j,v}
		      \fmf{dashes,label=$h_i$}{v,k}
		  \end{fmfgraph*}
  \end{fmffile}&
  \begin{fmffile}{v5}
		  \begin{fmfgraph*}(80,50)
		      \fmfleft{i,j}
	        \fmfright{k,l}	            
				\fmf{dashes,label=$A$,l.s=right,tension=1,l.dist=1.3}{i,v1}
		    \fmf{dashes,label=$A$,l.s=left,tension=1,l.dist=1.3}{j,v1}
		    \fmf{dashes,label=$h_i$,l.s=left,tension=1,l.dist=0.3}{k,v1}
		    \fmf{dashes,label=$h_j$,l.s=left,tension=1,l.dist=0.3}{v1,l}
	    \end{fmfgraph*}
\end{fmffile}
\\\hline
\multicolumn{4}{|c|}{\hspace*{-2.5cm} \vbox{\vspace{-0.1cm}\begin{equation*}
\begin{split}
 &i[\kappa v (R_{1i} R_{2,j} R_{2k} + R_{2,i} R_{1j} R_{2k} + R_{2,i} R_{2,j} R_{1k}) \\ 
 &+ \kappa \vs (R_{2,i} R_{1j} R_{1k} + R_{1i} R_{2,j} R_{1k} + R_{1i} R_{1j} R_{2k})\\
 &+ 6\lambda v ( R_{1i} R_{1j} R_{1k} ) + 6 \lambda_s \vs (R_{2,i} R_{2,j} R_{2k})]
 \end{split}
\end{equation*}\vspace{-0.3cm}}\hspace*{-2.5cm}} & 
 \begin{fmffile}{hihjhk}
	        \begin{fmfgraph*}(60,50)
	            \fmfleft{i,j}
	            \fmfright{k}
	      \fmf{dashes,label=$h_i$}{i,v1}
		    \fmf{dashes,label=$h_j$}{j,v1}
		    \fmf{dashes,label=$h_k$}{k,v1}
	        \end{fmfgraph*}
\end{fmffile}
\\\hline
\end{tabular}
\end{center}
\caption{Vertices relevant for the calculation of annihilation cross-section in the scalar DM model.}
\label{SDM_ver}
\end{table}

\subsection{Dark Matter Direct Detections}
\label{DD_dd_sdm}
It is interesting to note that the DM direct detection signals are naturally suppressed in the scalar DM model. It turns out that in the limit of zero DM velocity the tree-level amplitude for DM-nucleon scattering vanishes. The most relevant interaction term in this context is the $AAh_i$ vertex. From the potential Eq.~(\ref{pot_sdm}), one can easily derive the following DM triple-scalar couplings:
\begin{eqnarray}
V \supset \frac{A^2}{2} (2\lambda_S v_S \phi_S + \kappa v \phi_H) = \frac{A^2}{2v_S} (\sin\alpha\, m_1^2 h_1 + \cos\alpha\, m_2^2 h_2)\,, 
\label{GGH}
\end{eqnarray}
where we have used the relations Eqs.~(\ref{mixing_s}) and (\ref{relation_s}), and the corresponding Feynman rules are presented in tab.\ref{SDM_ver}. 
With these interaction terms, we can write down the corresponding amplitude for the spin-independent DM nuclear recoils as follows:
\begin{eqnarray}
i{\cal M} &=& - i \frac{\sin 2\alpha f_N m_N}{2v v_S} \left( \frac{m_1^2}{q^2 - m_1^2} - \frac{m_2^2}{q^2 - m_2^2} \right) \bar{u}_N(p_4) u_N(p_2) \nonumber\\
&\approx & -i \frac{\sin 2\alpha f_N m_N}{2v v_S} \left(\frac{m_1^2 - m_2^2}{m_1^2 m_2^2}\right) q^2 \bar{u}_N (p_4) u_N(p_2)\, ,
\end{eqnarray}
where $q^2$ represents the DM momentum transfer when it scatters with nucleons, and $m_N$ and $f_N \approx 0.3$ denote the nucleon mass and its coupling to the SM Higgs. In the limit of zero momentum transfer, $q^2 \to 0$, the above amplitude vanishes. This behaviour is a consequence of the fact that the Goldstone-Higgs coupling is proportional the Higgs mass squared. In the appendices 
we explain in a more general context when are the coupling of the form of (\ref{GGH}), i.e., $\propto m_i^2$. It is interesting to note~\cite{Liu:2017lpo} that a similar cancellation also exists in a DM model with a vector gauge boson mediator which communicates with the SM sector only through kinetic mixings with the SM neutral gauge bosons. As shown in Ref.~\cite{Liu:2017lpo}, it is even more remarkable that the cancellation in the vector mediator case does not demand to choose some specific soft breaking terms in the scalar potential, as in the present SDM.

It is shown in Ref.~\cite{Gross:2017dan} that the leading-order DM-nuclear recoil cross-section arises at one-loop order, which is estimated as follows by assuming the one-loop functions to be of ${\cal O}(1)$
\begin{eqnarray}\label{sigAN}
\sigma_{AN} \approx \left\{\begin{array}{cc}
\frac{\sin^2\alpha}{64\pi^5} \frac{m_N^4 f_N^2}{m_1^4 v^2} \frac{m_2^8}{m_A^2 v_S^6}\,, & m_A \geq m_{2} \\
\frac{\sin^2\alpha}{64\pi^5} \frac{m_N^4 f_N^2}{m_1^4 v^2} \frac{m_2^4 m_A^2}{v_S^6}\,, & m_A < m_{2}
\end{array}\right.\, .
\end{eqnarray}
The above result is a conservative estimate of the upper limit for the one-loop $A$-nucleon scattering cross-section. It turns out to be of ${\cal O}(10^{-49}~{\rm cm}^2)$ for $\sin\alpha = 0.1$, $m_2 = 300$~GeV and $m_A\sim 1~{\rm TeV}$, which is much lower than the current XENON1T limits of ${\cal O}(10^{-47}~{\rm cm}^2)$. Therefore, we expect that the DM direct searches will not impose any relevant constraints on the scalar DM model. In the following, we will use Eq.~(\ref{sigAN}) to perform the scan which indeed confirms this expectation.

\subsection{Higgs-boson invisable decays: \boldmath{$h_1\to AA$}}
\label{inv_dec_sdm}
One strong constraint for DM models comes from invisible decays of the SM-like Higgs boson, the corresponding branching  ratio should be less than $24\%$~\cite{Patrignani:2016xqp}. In the present scalar DM model with $m_A < m_1/2$, the SM Higgs boson decays invisibly into the stable pseudoscalar DM $A$, $h_1\to AA$, with the decay width given by
\begin{eqnarray}\label{widAA}
\Gamma (h_1 \to AA) = \frac{1}{32\pi} \frac{m_1^2 \sin\alpha^2}{v_S^2} \sqrt{m_1^2-4m_A^2}\,.
\end{eqnarray}

\section{Vector Dark Matter}
\label{VDM}

The model that we want to compare with the SDM is the popular vector DM (VDM) model~\cite{Hambye:2008bq,Lebedev:2011iq,Farzan:2012hh,Baek:2012se,Baek:2014jga,Duch:2015jta} 
that is an extension of the SM by an additional $\uone$ gauge symmetry and a complex scalar field 
$S$, whose vev generates a mass for this $U(1)$'s vector field. The quantum numbers of the scalar field are 
\beq
S  =   (0,{\bf 1},{\bf 1},1) \ \  \text{under}  \ \  U(1)_Y\times SU(2)_L \times SU(3)_c \times \uone.
\eeq
None of the SM fields are charged under the extra gauge group.  In order to ensure stability of the new vector boson
a $\z2$ symmetry is assumed to forbid $U(1)$-kinetic mixing between $\uone$ and  $U(1)_Y$. The extra gauge boson $A_\mu$  and 
the scalar field $S$  transform under the $\z2$ as follows
\beq
A^{\mu}_X \rightarrow -A^{\mu}_X \ , \ S\rightarrow S^*,  \ \text{where} \  S=\phi e^{i\sigma}, \ \ {\rm so} \ \ \phi\rightarrow \phi,  \ \  \sigma\rightarrow -\sigma.
\eeq
All other fields are neutral under the $\z2$. 

At leading order the vector bosons masses are given by:
\beq
M_W=\inv2 g v, \ \ \ \ M_Z = \frac{1}{2}\sqrt{g^2+g'^2} v \ \ \ \text{and} \ \ \   m_X = \gx \vs,
\eeq
where $g$ and $g'$ are the $SU(2)$ and $U(1)$ gauge couplings, while $v$ and $\vs$ are $H$ and $S$ vev's: $(\langle H \rangle,\langle S \rangle)=\frac{1}{\sqrt{2}}(v,\vs)$.
The scalar potential for this model is given by
\beq
V= -\mu^2_H|H|^2 +\lambda_H |H|^4 -\mu^2_S|S|^2 +\lambda_S |S|^4 +\kappa |S|^2|H|^2 .
\eeq
It will also be useful to define, for future reference, the parameter $\lambda_{SM}\equiv m_1^2/(2 v^2)=0.13$, 
where $m_1\equiv 125.09 \gev$. 

The requirement of positivity for the potential implies the following constraints that we impose in all further discussions:
\beq
\lambda_H > 0, \ \ \lambda_S >0, \ \ \kappa > -2 \sqrt{\lambda_H \lambda_S}.
\eeq

It is easy to find the minimization conditions for the scalar fields 
(without losing generality one can assume $v,\vs>0$):
\beq
(2\lambda_H v^2 + \kappa \vs^2 - 2\mu^2_H) v = 0\ \  \text{and} \ \  (\kappa v^2 + 2\lambda_S \vs^2 - 2\mu^2_S)\vs = 0
\label{min_con}
\eeq
If $\mu_{H,S}^2<0$ the global minimum at $(0,0)$ is the only extremum. For $\mu_{H,S}^2>0$ the point $(0,0)$ is a local maximum of the potential,
in this case $(0,\frac{\mu_S}{\sqrt{\lambda_S}})$ and $(\frac{\mu_H}{\sqrt{\lambda_H}},0)$ are global minima if 
$\kappa^2>4\lambda_H\lambda_S$, otherwise they are saddle points and the global minima are determined by
\beq
v^2=\frac{4 \lambda_S \mu^2_H - 2\kappa \mu^2_S }{4\lambda_H\lambda_S-\kappa^2},\ \ \vs^2=\frac{4 \lambda_H \mu^2_S - 2\kappa \mu^2_H }{4\lambda_H\lambda_S-\kappa^2}.
\eeq
For the VDM model only the latter case is relevant, since both vevs need to be non-zero to give rise to the masses of the SM fields and of the dark vector boson. 
Both scalar fields can be expanded around corresponding vev's as follows
\beq
S=\frac{1}{\sqrt{2}}(\vs+ \phi_S +i\sigma_S)  \ \ , \ \  H^0= \frac{1}{\sqrt{2}}(v + \phi_H+ i\sigma_H)   \ \  \text{where} \ \  H=\binom{H^+}{H^0}.
\eeq
The mass squared matrix $\mathcal{M}^2$ for the fluctuations $ \left(\phi_H, \phi_S\right)$ reads
\begin{equation} 
\mathcal{M}^2 = \left( 
\begin{array}{cc}
2 \lambda_H v^2  & \kappa v \vs \\ 
 \kappa v \vs &2 \lambda_S v^2_x 
\end{array} 
\right).
\label{massmatrix}
 \end{equation} 
where the similarity to the mass matrix~\ref{massv1} in the SDM model is obvious. This mass matrix $\mathcal{M}^2$ can be diagonalised by the orthogonal rotation $R$ exactly as in
\ref{mix} for the SDM. Note that here we adopt a convention such  $h_1$ is the observed Higgs particle.

There are 5 real parameters in the potential: $\mu_H$, $\mu_S$, $\lamh$, $\lams$ and $\kappa$. Adopting the minimization conditions
(\ref{min_con}) $\mu_H$, $\mu_S$ can be replaced by $v$ and $\vs$. 
Eventually there are 4~independent unknown parameters in the model and a convenient choice in this project is 
$\vs, \sin\alpha, m_2$ and $m_X$, which matches the choice made for the SDM model. 
The parameters of the potential can be written as a function of the above set as:
\bea
\lamh &=& \lamsm+\sin^2\alpha \frac{\mtwo^2-\mone^2}{2v^2} \\
\kappa^2 &=& 4 (\lamh-\lamsm) \frac{\lams \vs^2-\lamsm v^2}{\vs^2} \\
\lams &=& \frac{2\kappa^2}{\sin^2 2\alpha}\frac{v^2}{\mtwo^2-\mone^2}\left( \frac{\mtwo^2}{\mtwo^2-\mone^2}-\sin^2\alpha  \right).
\eea

The extra vertices (besides those shown in tab.~\ref{SDM_ver}) needed for further calculations are collected in tab.~\ref{extra_ver}. 

\begin{table}[h]
\begin{center}
\te
\begin{tabular}{|c|c|}
\hline
& \vspace{-0.5cm} 
\\
$ i2m_X\gx R_{2,i}$
& $i2\gx^2 R_{2,i}R_{2,j}$
\\
& \vspace{-0.5cm} 
\\
\hline
& \vspace{-0.5cm} 
\\  
\begin{fmffile}{ZpZphi}
	        \begin{fmfgraph*}(60,50)
	            \fmfleft{i,j}
	            \fmfright{k}
	      \fmf{boson,label=$X$,l.s=right}{i,v}
		    \fmf{boson,label=$X$,l.s=left}{j,v}
		    \fmf{dashes,label=$h_i$}{v,k}
	        \end{fmfgraph*}
\end{fmffile}&
  \begin{fmffile}{hhZpZp}
		  \begin{fmfgraph*}(80,50)
		      \fmfleft{i,j}
	            \fmfright{k,l}
	      \fmf{boson,label=$X$,l.s=right,l.dist=0.7}{i,v1}
		    \fmf{boson,label=$X$,l.s=left,l.dist=3.7}{j,v1}
		    \fmf{dashes,label=$h_i$,l.s=left,l.dist=0.4}{k,v1}
		    \fmf{dashes,label=$h_j$,l.s=left,l.dist=0.4}{v1,l}
	        \end{fmfgraph*}
\end{fmffile}
\\\hline
\end{tabular}
\end{center}
\caption{The extra vertices relevant for the calculation of annihilation and scattering cross-sections in the vector DM model.}
\label{extra_ver}
\end{table}

\subsection{Dark Matter Direct Detection}
\label{DD_dd_vdm}

The VDM model is constrained by the direct detection experiments. The spin-independent $XN$ scattering cross-section is given by~\cite{Duch:2017khv}
\begin{eqnarray}\label{sigXN}
\sigma_{XN} = \frac{\sin^2{2\alpha}}{4\pi} \frac{(m_1^2 -m_2^2)^2}{m_1^4 m_2^4} \frac{f_N^2 \mu_{XN}^2 m_X^2 m_N^2}{v^2 v_S^2}\,,
\end{eqnarray}
where $\mu_{XN} \equiv m_X m_N/(m_X+m_N)$ is the reduced mass in the DM-nucleon system. Note that compared with the pseudoscalar DM case in Eq.~(\ref{sigAN}), it is clear that there is no suppression due to additional powers of relative DM velocity, thus we expect that the DM direct detection to results in a strong constraint to the present VDM model. 


\subsection{Higg-boson invisable decays: \boldmath{$h_1\to XX$}}
\label{inv_dec_vdm}

When the VDM mass is smaller than half of the SM-like Higgs boson $h_1$, $m_X < m_1/2$, the Higgs invisible decay provides another constraint on the VDM scenario. In the present model, the width for invisible decays is provided by the process $h_1 \to XX$ and can be expressed as follows~\cite{Duch:2017khv}
\begin{eqnarray}\label{widXX}
\Gamma(h_1 \to XX) = \frac{g_X^2 \sin^2\alpha}{8\pi} \sqrt{m_1^2 -4m_X^2} \frac{m_X^2}{m_1^2} \left[2+ \frac{(m_1^2 - 2m_X^2)^2}{4m_X^4} \right]\,.
\end{eqnarray} 

\section{Disentangling the scalar and vector DM models at future linear \boldmath{$e^+e^-$} colliders}
\label{coll_tests}
\begin{figure}[!h]
		\centering
		\includegraphics[width=0.4\textwidth]{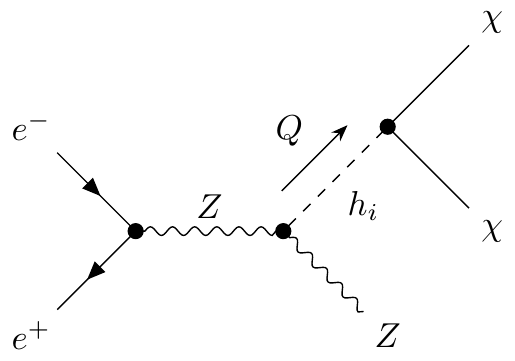}
		\vspace*{0.5cm}
		\caption{Feynman diagram for considered channel of DM production. $\chi$ denotes the dark particle ($\chi=A,X$).}
		\label{fig:eeHZ}
	\end{figure}

The DM Higgs portal models can be tested by collider experiments \cite{Birkedal:2004xn,Feng:2005gj,Konar:2009ae,Goodman:2010ku}. The different DM scenarios were discussed using the effective operator approach \cite{Fox:2011pm,Kumar:2015wya}, simplified models \cite{Alves:2011wf,Abercrombie:2015wmb,Buckley:2014fba,Abdallah:2015ter,Abe:2018bpo} or other simple renormalizable models respecting unitarity and gauge-invariance\cite{Chen:2015dea,Baek:2015lna,Ko:2016ybp}. An especially promising tool to probe the DM models discussed in this paper are future $e^+e^-$ colliders \cite{Baltz:2006fm}. In particular, they allow for the copious production of DM states associated with a $Z$~boson, what is referred to as Higgsstrahlung process or mono-$Z$ emision \cite{Dreiner:2012xm,Yu:2014ula,Neng:2014mga,Ko:2016xwd,Liu:2017lpo,Rawat:2017fak,Kamon:2017yfx} (see the diagram in Fig.~\ref{fig:eeHZ}).
We assume that the energy of the $Z$ boson can be reconstructed from data, therefore allowing for the determination of the missing energy, corresponding to the dark particles. 
The number of events observed for a given energy bin $(E_Z,E_Z+\Delta E_Z)$ allows to measure the value of the differential cross-section, $\frac{d\sigma}{dE_Z}(E_Z)$, which is given by the following formula:
	\begin{align}\begin{aligned}\label{eq:dsigma}
	\frac{d\sigma}{dE_Z}(E_Z)=
	&f(s,E_Z)\cdot
	\frac{\left(\frac{\sin{2\alpha}}{v_S}\right)^2\cdot\sqrt{1-4\frac{\mdm^2}{Q^2}}\cdot (m_1^2-m_2^2)^2\cdot Q^4}
	{\left[(Q^2-m_1^2)^2+(m_1\Gamma_1)^2\right]\left[(Q^2-m_2^2)^2+(m_2\Gamma_2)^2\right]}\times \\
	&\times
		\begin{cases}
		1 & (\text{SDM})\\
		1 - 4\frac{m_{DM}^2}{Q^2} + 12 \left(\frac{m_{DM}^2}{Q^2}\right)^2 & (\text{VDM})
		\end{cases},
	\end{aligned}\end{align}
where 
	\begin{gather}
	f(s,E_Z)\equiv\frac{(1-P_+P_-)(g_v^2+g_a^2)+2g_vg_a(P_+-P_-)}{12\cdot(2\pi)^3}\sqrt{E_Z^2-m_Z^2}\left(2 m_Z^2+E_Z^2\right)
	\left(\frac{g^2}{\cos\theta_W^2}\frac{1}{s-m_Z^2}\right)^2,\\
	Q^2=Q^2(s,E_Z)\equiv s-2E_Z\sqrt{s}+m_Z^2.
	\end{gather}
Here $g_v=\frac{1}{2}(1-4\sin^2\theta_W)$ and $g_a=\frac{1}{2}$ are the vector and axial couplings between electrons and the $Z$ boson, $g$ is the weak coupling constant, $m_Z$ is mass of the $Z$ boson and $\theta_W$ denotes the Weinberg angle. $P_+$ and $P_-$ denote the polarisation (defined as in \cite{Fujii:2018mli}) of the positron and electron beam, respectively. Employing polarised beams can help to reduce the SM background (see section \ref{sec:error}). The mass of the dark particle is denoted by $m_{DM}$ (it is $m_A$ for the SDM and $m_X$ for the VDM) and $Q^2$ is the squared four-momentum of the decaying Higgs particle. $\Gamma_1$ and $\Gamma_2$ are the total (including SM as well as dark channels) decay widths of $h_1$ and $h_2$, respectively, which must be calculated within each model as follows
\begin{align}\label{eq:width}
	\Gamma_i=\Gamma_i^\text{SM}+\frac{R_{2i}^2}{32\pi}\frac{m_i^3}{v_S^2}\sqrt{1-\frac{4\mdm^2}{m_i^2}} \cdot
		\begin{cases}
		1 & (\text{SDM})\\
		1 - 4\frac{m_{DM}^2}{m_i^2} + 12 \left(\frac{m_{DM}^2}{m_i^2}\right)^2 & (\text{VDM})
		\end{cases},
\end{align}
where $\Gamma_i^\text{SM}$ is the width of $h_i$ into SM final states. Note that the widths in (\ref{eq:dsigma}) were dropped in the numerator as they are higher order terms in the perturbation expansion.
Since $Q^2 \geq 4 \mdm^2$, the following important inequality holds
\beq
\text{max}\left[\frac34, f(x_\text{min})\right]\geq 1 - 4\frac{m_{DM}^2}{Q^2} + 12 \left(\frac{m_{DM}^2}{Q^2}\right)^2 \geq
\begin{cases}
\frac23     & \text{if }x_{\min}<\frac23\\
f(x_{\min}) & \text{if } \frac23<x_{\min}<1
\end{cases}\, ,
\label{fun-limit}
\eeq
where
\beq 
f(x)\equiv 1-x+\frac34 x^2 \lsp \text{and} \lsp x_\text{min}\equiv \left(\frac{2\mdm}{\sqrt{s}-m_Z}\right)^2\, .
\eeq
Therefore from (\ref{eq:dsigma}) we obtain the following solid prediction for the ratio of differential cross-sections for SDM and VDM:
\beq
\left\{ \text{max}\left[\frac34, f(x_\text{min})\right]\right\}^{-1} \; \lsim \; \frac{\frac{d\sigma_\sdm}{dE_Z}}{\frac{d\sigma_\vdm}{dE_Z}} \; \lsim \; 
\begin{cases}
\frac32               & \text{if }x_{\min}<\frac23\\
[f(x_{\min})]^{-1} & \text{if } \frac23<x_{\min}<1
\end{cases} \, ,
\label{r-limit}
\eeq
where it was assumed that the decay widths of $h_{1,2}$ are similar in both models.
For cases adopted in this section $\text{max}\left[3/4, f(x_\text{min})\right] \simeq 1$ therefore the lhs of inequality 
\ref{r-limit} is very close to $1$ while the rhs is $3/2$.
As a consequence of the above inequality, the total number of events predicted for the SDM model must be greater than for the VDM.
The maximal deviation of the ratio of the distributions (\ref{eq:dsigma}) from 1 corresponds to $Q^2=6m_{DM}^2$. Hence, it is easy to find that
the distance $\delta$ between the energy $E_Z$ corresponding to the maximal deviation and the location of the $i$-th pole is
\beq
\delta=\frac{m_i^2-6m_{DM}^2}{2 \sqrt{s}} \, .
\label{delta}
\eeq
In turn, this means that the regions where the large ratio of the distributions (\ref{eq:dsigma}) occur, are in the vicinity of a resonance (where the number of events is expected to be large). 
In particular, if $m_i^2=6 m_{DM}^2$ the maximal deviation ($50\%$) appears exactly at the $i$-th pole.

\begin{figure}[!h]
		\centering
		\includegraphics[width=0.75\textwidth]{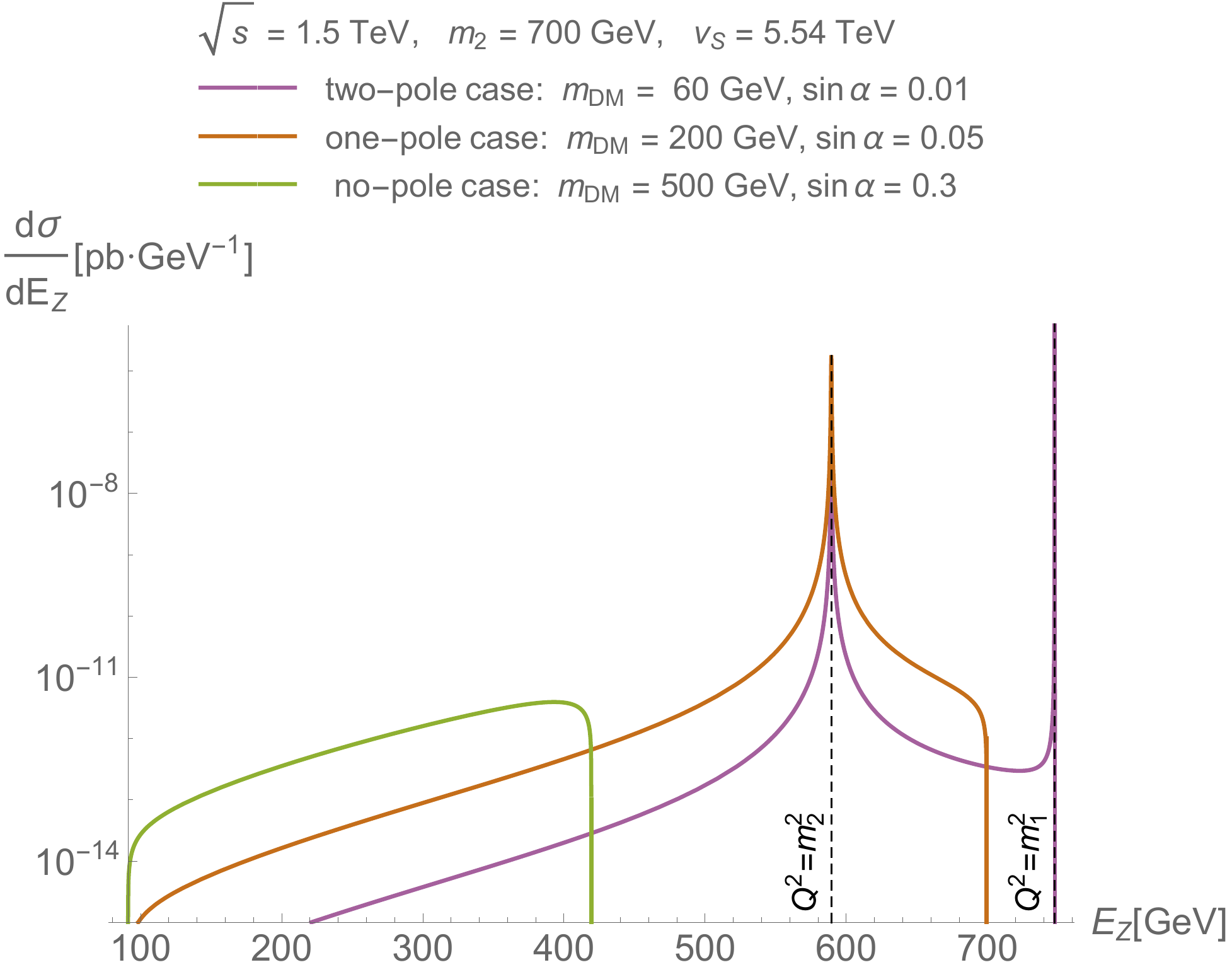}
		\vspace*{0.5cm}
		\caption{An exemplary plot of $\frac{d\sigma}{dE_Z}$ function for the SDM model. Different curves correspond to different cases: for the purple one, $2\cdot m_{DM}<m_1,m_2$; for the brown $m_1<2\cdot m_{DM}<m_2$; and for the green $m_1,m_2<2\cdot m_{DM}$.}
		\label{fig:dsigma0}
	\end{figure}

An exemplary plot of $\frac{d\sigma}{dE_Z}$ is presented in figure \ref{fig:dsigma0}.
The maximal value of $E_Z$ for this process is 
	\begin{align}
	E_{\max}=\frac{s-4\mdm^2+m_Z^2}{2\sqrt{s}},
	\end{align}
what corresponds to $Q^2=4\mdm^2$. If $E_Z$ was higher, there would not be sufficient energy to produce the dark particles. Note that this threshold is clearly visible on the plot and we 
therefore assume that the mass of dark particles can be read from data. 

The poles, $Q^2=m_i^2$, correspond to $h_i$ being on-shell. Therefore, the $i$-th pole is present if
	\begin{align}
	2\cdot m_{DM}<m_i<\sqrt{s}-m_Z.
	\end{align}
In this case energy of $Z$ boson is
	\begin{align}\label{eq:poleE}
	E_Z(Q^2=m_i^2)=E_i\equiv\frac{s-m_i^2+m_Z^2}{2\sqrt{s}}.
	\end{align}
which in turn means that the mass of $h_2$ can be read from the position of the $h_2$ pole. If the $h_2$ pole is not present, 
$m_2$ has to be determined by an independent measurement. 

Recently, two papers~\cite{Ko:2016xwd,Kamon:2017yfx} have discussed similar issues as the one described in this section. Their authors have considered the possibility to disentangle vector, scalar and fermion DM at $e^+e^-$ colliders. The vector model they adopted is the same as the one discussed here. However, for the scalar DM they used a minimal model with an extension by a real singlet, not by a complex one with softly broken global $U(1)$, which has been adopted here.
Note that the scalar model considered here and the one adopted in \cite{Ko:2016xwd,Kamon:2017yfx} are, in fact, very different.
There, the coupling between DM and the mediator (the SM Higgs boson) is just given by the Higgs portal coupling (and the SM vev) and is independent of the mediator mass. In contrast, in the model discussed here, the DM is a pseudo-Goldstone boson of the spontaneously broken $U(1)$ symmetry. It is easy to see that in the limit of restored symmetry, i.e. when $\mu^2\to 0$, the DM $A$ becomes a massless genuine Goldstone boson. As it is shown in the appendices, in our case, with the $U(1)$ broken softly by the quadratic term $\mu^2(S^2+S^{*\,2})$, the coupling between the DM and the mediator ($h_i$) is proportional to the mediator mass squared $m_i^2/v_s$. Note that in the VDM model the corresponding degree of freedom is a would-be Goldstone $G_X$ boson which becomes the longitudinal component of the massive vector DM $X$. Nevertheless, when one compares the ILC potential for those two versions of scalar DM versus VDM, it turns out that our conclusions are slightly less optimistic than those published in \cite{Ko:2016xwd,Kamon:2017yfx}.

In the following subsections we present a comparison of both DM models in a few typical cases for a $\sqrt{s}=1.5\tev$ collider.

\begin{figure}[!h]
		\centering
		\includegraphics[width=\textwidth]{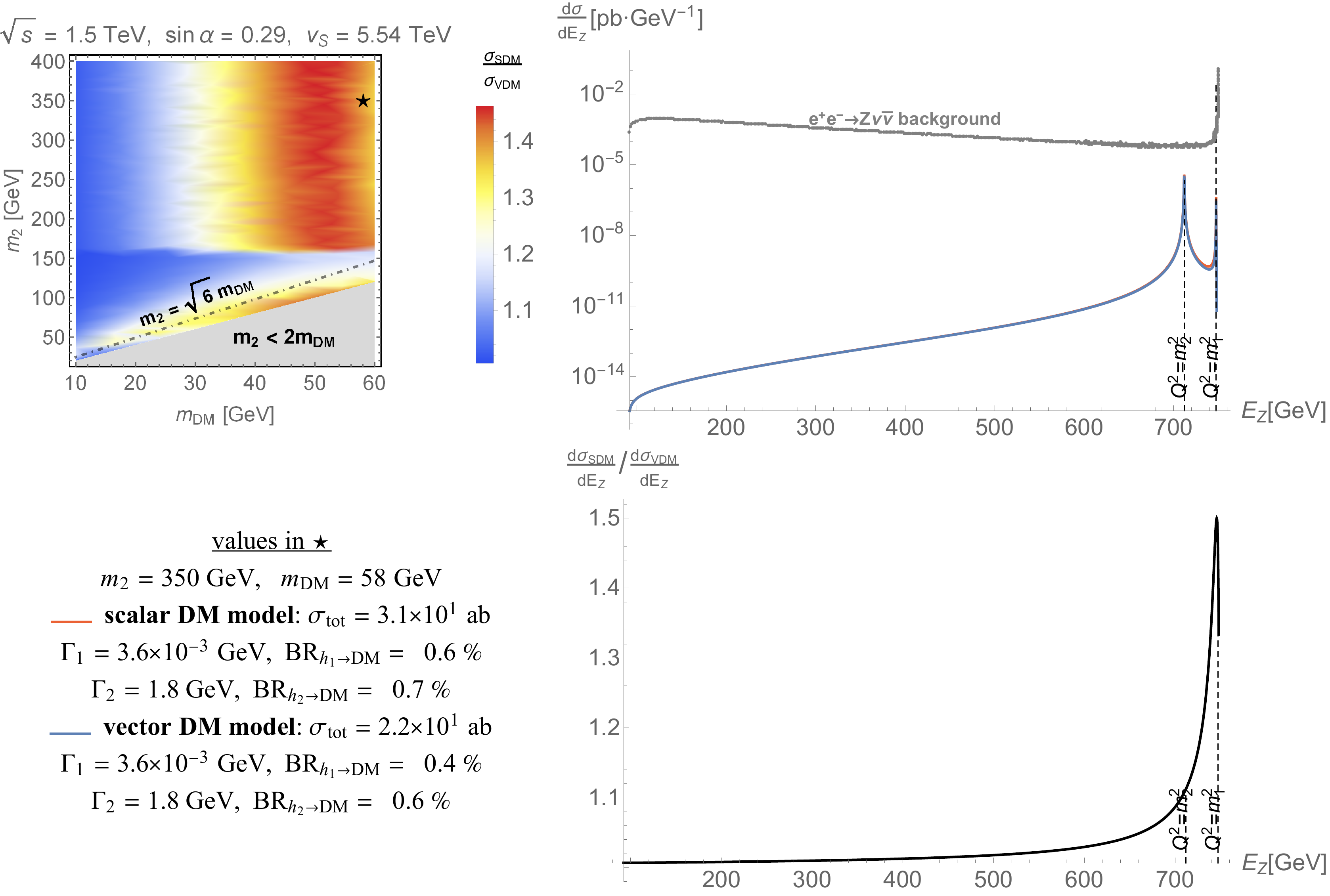}
		\vspace*{0.5cm}
		\caption{Comparison of cross-sections for the $e^+e^-\to Zh_i(\chi\chi)$ process ($\chi=A,X$) for the SDM and for the VDM, in the two-pole case: $2\cdot m_{DM}<m_1,m_2$. The upper right panel shows $d\sigma/d E_Z$ for both models while the lower one shows the ratio of the distributions between the SDM and the VDM. The parameters chosen for the plot in the right panels are specified in the lower left corner and above the upper left panel. The chosen values for $(m_{DM},m_2)$ correspond to the point denoted by the star in the left upper panel. The colour bar shows the value of the ratio of total cross-sections for $e^+e^-\to Zh_i(\chi\chi)$. The thick gray line on the upper right panel represents the SM background for our process (see section~\ref{sec:error}). Polarizations of the beams are $(P_+,P_-)=(-30\%,80\%)$.}
		\label{fig:h2PoleRatio1}
\end{figure}

\subsection{Two-pole case}

In this section we assume that both poles are present. As already mentioned $m_2$ and $m_{DM}$ could be determined by the location of the $h_2$ resonance and by the endpoint of the distribution. We assume
 that $\sin\alpha$ and $v_S$ are known (deduced from some independent measurements), so that we can compare the two models at the same points in the parameter space.

Fig. \ref{fig:h2PoleRatio1} presents contours of the ratio of total cross-sections $\sigma_{SDM}/\sigma_{VDM}$ in the $(m_{DM},m_2)$ space. The structure expected from (\ref{delta}) is visible, we observe the enhancement of the ratio for $m_{DM}\simeq m_1/\sqrt{6}\simeq 51\gev$ and also for $m_2\simeq \sqrt{6}\; m_{DM}$. In those regions $\sigma_{SDM}/\sigma_{VDM}$ reaches its maximal value $\sim 1.5$. The right panels show that, for the parameters chosen there, maximal enhancement of $d\sigma/d E_Z$ is observed near the resonance $Q^2=m_1^2$ and therefore a substantial value for the ratio of the total cross-sections ($\sim 1.40$) could be reached. The point in the parameter space adopted in the right panel satisfies all the experimental and theoretical constraints considered here. 
The region for which a two-pole scenario is not possible is marked in gray.
\subsection{One-pole case}

\begin{figure}[!h]
		\centering
		\includegraphics[width=\textwidth]{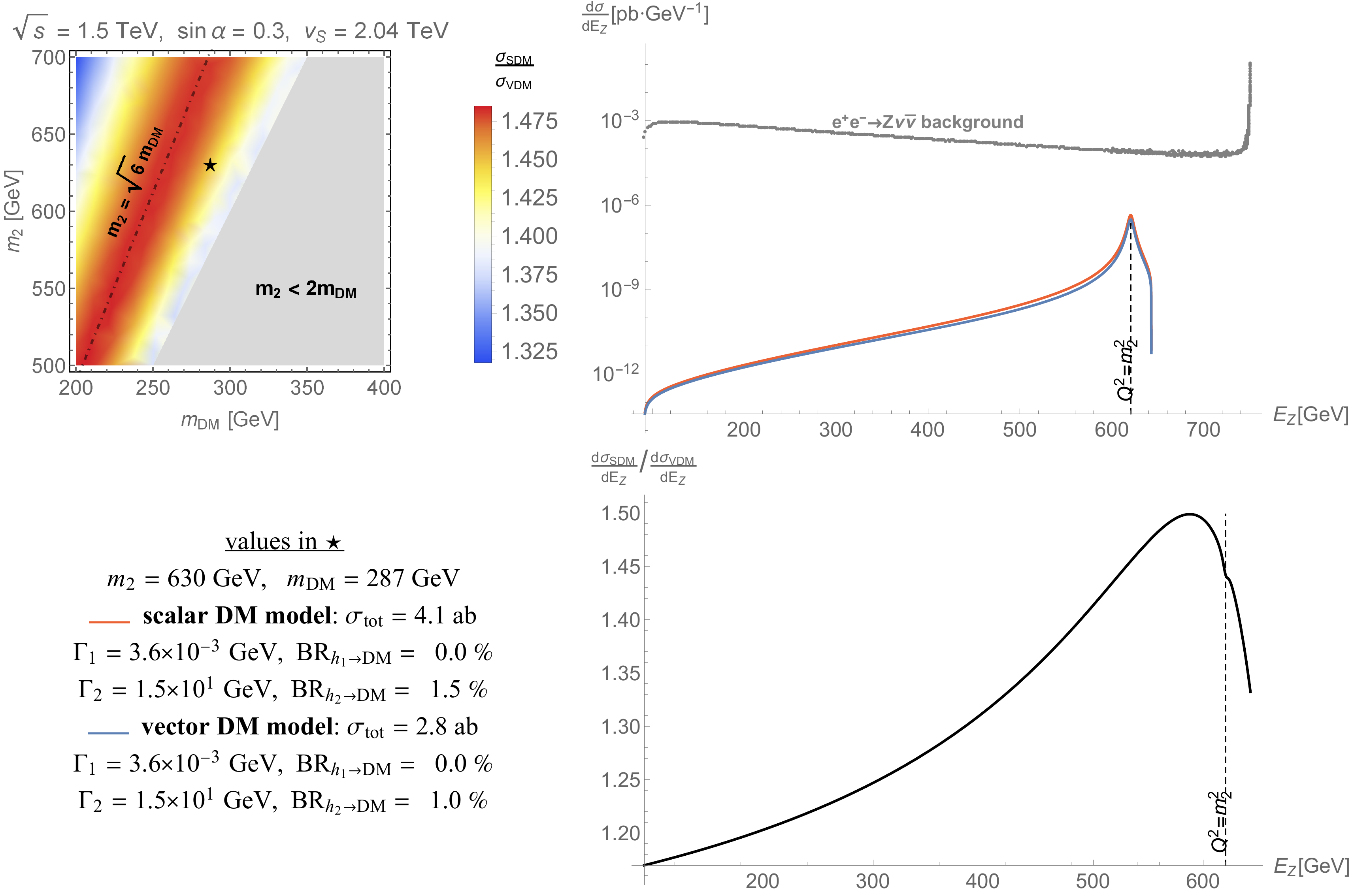}
		\vspace*{0.5cm}
		\caption{As in fig.~\ref{fig:h2PoleRatio1}, however for the one-pole case, i.e. for $m_1<2\cdot m_{DM}<m_2$.}
		\label{fig:h2PoleRatio2}
	\end{figure}

In this scenario we assume that $m_1<2\cdot m_{DM}<m_2$, therefore only one of the poles could be observed. 
Fig. \ref{fig:h2PoleRatio2} shows the distribution functions and the ratio of the total cross-sections in this case. We also show the contour plot of the ratio of total cross-sections $\sigma_{SDM}/\sigma_{VDM}$ in the $(m_{DM},m_2)$ space. Since $m_1<2\cdot m_{DM}$ only the $h_2$ resonance appears. Again, for $m_2\simeq \sqrt{6}\; m_{DM}$ the ratio of total cross-sections is observed 
with maximal value close to $1.5$, i.e. maximal possible enhancement. There is only one enhancement band present in this case and the point in the parameter space adopted in the right panels satisfy all the experimental and theoretical constraints considered here. The point has been chosen such that the maximal ratio of the differential cross-sections is observed near the resonance, so that the ratio of the total cross-sections can reach $\sim 1.45$. 

\subsection{No-pole case}

\begin{figure}[!h]
		\centering
		\includegraphics[width=\textwidth]{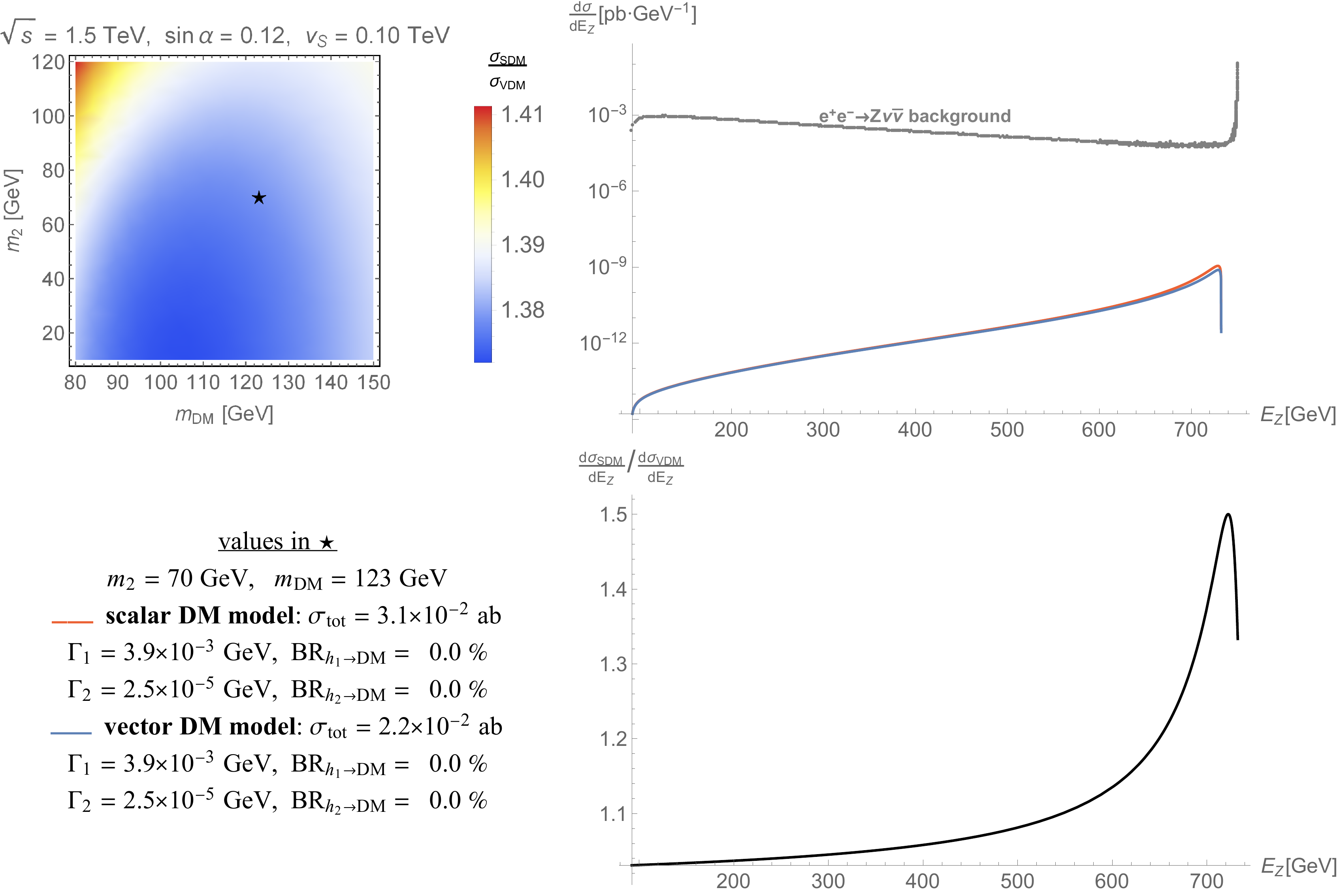}
		\vspace*{0.5cm}
		\caption{As in fig.~\ref{fig:h2PoleRatio1}, however for the no-pole case: $m_1,m_2<2 \cdot m_{DM}$.}
		\label{fig:h2PoleRatio3}
	\end{figure}

In this case no pole is present since both Higgs particles are lighter than $2 \cdot m_{DM}$. Again we adopt a similar strategy to illustrate this case. The difference is that since there is no pole present the mechanism to amplify the ratio of $\sigma_{SDM}/\sigma_{VDM}$ does not work. As a result, the contour plots for the ratio of the total cross-sections show only very mild enhancement this time. These results are shown in Fig. \ref{fig:h2PoleRatio3}.

\subsection{Expected statistical error and SM background}\label{sec:error}

\begin{figure}[!h]
		\centering
		\includegraphics[width=\textwidth]{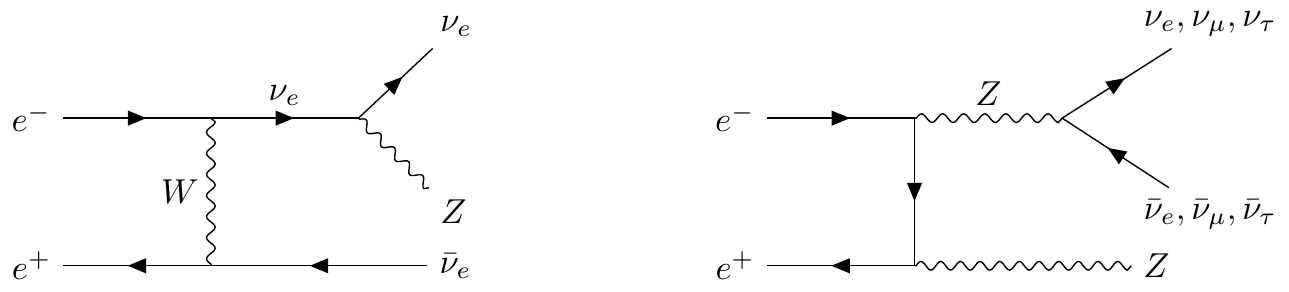}
		\vspace*{0.5cm}
		\caption{Exemplary diagrams of the Standard Model background processes. Neutrinos contribute to missing energy and can therefore mimic dark particles. The background cross-section could be reduced by polarizing the initial $e^+$ and $e^-$ beams. }
		\label{fig:bckgrnd}
\end{figure}

The expected statistical error for measurements of the total cross-section is equal to
	\begin{align}
	\Delta\sigma=\sqrt{\frac{\sigma_\text{sig}+\sigma_\text{bg}}{\eta\displaystyle\int\!\!\mathscr{L}dt}},
	\end{align}
where $\eta$ stands for the efficiency of detectors, $\displaystyle\int\!\!\mathscr{L}dt$ is the luminosity of the collider integrated over the whole data collection period for a given $\sqrt{s}$, 
$\sigma_\text{sig}$ is the total cross-section for the signal and $\sigma_\text{bg}$ is the background cross-section. Following \cite{Dannheim:2012rn,CEPCStudyGroup:2018ghi}, we assume that
	\begin{align}
	\eta&\approx 1,
	&\int\!\!\mathscr{L}dt\Big|_{\text{CLIC},\sqrt{s}=1.5\tev}
	&\approx 1500\text{ fb}^{-1},
	&\int\!\!\mathscr{L}dt\Big|_{\text{CEPC},\sqrt{s}=240\gev}
	&\approx 3000\text{ fb}^{-1}.
	\end{align}
In order to decide whether the two models will be experimentally distinguishable, we can compare the difference between the total cross-sections for SDM and VDM to its uncertainty 
	\begin{align}
	\Delta(\sigma_\sdm-\sigma_\vdm)=
	\sqrt{(\Delta\sigma_\sdm)^2+(\Delta\sigma_\vdm)^2}=
	\sqrt{\frac{(\sigma_\sdm+\sigma_\text{bg})+(\sigma_\vdm+\sigma_\text{bg})}
	{\eta\displaystyle\int\!\!\mathscr{L}dt}}.
	\end{align}
Production of dark particles can be mimicked by processes that produce missing energy in the form of SM neutrinos that escape detection. The simplest and most relevant example of such a process, with a similar experimental signature, that is the irreducible background, is $e^+e^-\to ZZ^*\to Z\nu\bar\nu$, where $\nu=\nu_e,\nu_\mu,\nu_\tau$ (see the exemplary Feynman diagrams in Fig.~\ref{fig:bckgrnd}). The background could be reduced to some extent by employing polaraised $e^+e^-$ beams. According to the current predictions for the ILC \cite{Adolphsen:2013kya}, polarisation of the beams should be possible at the level of $80\%$ for electrons and $30\%$ for positrons. The expected background, presented in Figs.~\ref{fig:h2PoleRatio1}-\ref{fig:h2PoleRatio3}, is calculated under that assumption using CalcHEP \cite{Belyaev:2012qa}. The background is a serious obstacle in the determination of parameters for DM particles. One can see that among our examples only in the two-pole case in the vicinity of the SM-like resonance $h_1$, the signal is comparable to the background. In order to increase the signal-to-background ratio, one should be collecting only events in a vicinity of a pole for one of the Higgs bosons, i.e. only events with $Z$ boson energy within a certain bin around $E_Z=E_i(\sqrt{s})\equiv(s-m_i^2+m_Z^2)/(2 \sqrt{s})$. It turns out that for $\sqrt{s}=1.5\tev$, $\mdm=44.5\gev$, $m_2=102\gev$, $\vs=5\tev$ and $\sin\alpha=0.31$ the separation between the cross-sections for SDM and VDM at the level of 1$\sigma$ could be obtained for a bin around $E_Z=E_1(1.5\tev)$ with the width $\sim 4.5\gev$. To estimate the minimal value of experimental uncertainty of $E_Z$ we assumed that energy of the Z boson is reconstructed from energy of jets that are produced, since the branching ratio of $Z$ into hadrons is almost $70\%$~\cite{PhysRevD.98.030001}. The jet energy can be measured in calorimeters with resolution $\sim 3\%$~\cite{CEPCStudyGroup:2018ghi,Linssen:2012hp}. Hence, the minimal size expected for the resolution of the $Z$ energy near the $h_1$ pole at, for instance, CLIC is $\sim 3\%\times E_Z|_{E_Z=E_1(1.5\tev)}=22.4\gev$, and so it seems extremely hard to disentangle the two scenarios for the adopted parameters at this collider. On the other hand, for $\sqrt{s}=240\gev$ expected for the CEPC and the same parameters, for the minimal bin size $\sim 3\%\times E_Z|_{E_Z=E_1(240\gev)}=3.1\gev$ the separation between the two cross-sections is at the level of 12$\sigma$. Therefore it is fair to conclude that there exist regions of parameters, where the two scenarios might be disentangled at future $e^+e^-$ colliders in resonance regions. However, without a detailed error and background analysis that takes into account all experimental details it is impossible to draw any solid final conclusions. 

\section{Numerical simulation}
\label{numericalsim}

The two models described in the previous sections were implemented in the \texttt{ScannerS}~\cite{Coimbra:2013qq, Costa:2015llh} code as model classes. 
The code takes as input any scalar potential that is a polynomial in the fields of order up to four and by considering the VEVs, mixing angle and physical masses 
as independent parameters, turns the problem of deriving the original potential parameters into a set of linear equations, with a very significant increase in 
speed of the scanning process (see~\cite{Coimbra:2013qq} for details). In the most general cases, the drawback of this method is that a given point is only 
verified to be a global minimum at the end of the procedure. However, because it is easy to obtain closed conditions for the global minimum for the
particular models under study, this problem is avoided. 
The code is equipped with a set of tools which allow to automatise the parameter scans and also with generic modules that allow to test local vacuum
stability and library interfaces to the constraints implemented for each model.  \texttt{ScannerS} is also interfaced with other high energy tools
that simplify the implementation of the constraints that will be described shortly. 

The ranges for the independent parameters are listed in Table~\ref{tab:par_ranges}. The ranges are the same for both models under study.

\begin{table}[!h]
	\centering
	\begin{tabular}{lr}  
		\toprule		
		Parameter & Range \\
		\midrule
		SM-Higgs - $m_1$ & 125.09 GeV \\
		Second Higgs - $m_2$ & [1,1000] GeV\\
		DM - $m_\text{DM}$ & [1,1000] GeV  \\
		Singlet VEV - $v_s$ & [1,$10^7$] GeV  \\
		Mixing angle - $\alpha$ & [$-\frac{\pi}{4}$,$\frac{\pi}{4}$] \\
		\bottomrule
	\end{tabular}
\vspace*{0.5cm}
	\caption{Independent parameters' range for both models.}
	\label{tab:par_ranges}
\end{table} 

The points generated using \texttt{ScannerS} have to be in agreement with the most relevant experimental and theoretical constraints.  The discovered
Higgs boson mass is taken to be $m_h = 125.09$ GeV from the ATLAS/CMS combination~\cite{Aad:2015zhl}. In these models the Higgs couplings
to remaining SM particles are all modified by the same factor. Therefore, the bound on the signal strength~\cite{Aad:2015zhl} is used to constrain
this parameter. The vacuum expectation value
of the Higgs doublet is fixed by the W-mass. The points generated have to comply with the following theoretical constraints: i) the potential has to be bounded from below;
ii) the vacuum is chosen so that the minimum is the global one and iii) perturbative unitarity holds. The first two constraints are implemented in the code 
while perturbative unitarity is imposed trough an internal numerical procedure that includes all possible two to two processes and that is available
in \texttt{ScannerS} for a generic model.
In these models new contributions to the radiative corrections of the massive gauge-boson self-energies, $\Pi_{WW} (q^2)$ and  $\Pi_{ZZ} (q^2)$
appear via the mixing between the neutral components of the doublet and the singlet. We use the variables S, T, U \cite{Peskin:1991sw} (expressions available in \cite{Grimus:2008nb} ) to guaranty that 
the models are in agreement with the electroweak precision measurements at the 2$\sigma$ level. 

The phenomenological constraints are imposed either via libraries in the code or with interfaces with other high energy codes. The collider bounds   
from LEP, Tevatron and the LHC are all encoded in~\texttt{HiggsBounds}~\cite{Bechtle:2008jh}. The program can be used to ensure agreement 
at 95\% confidence level exclusion limits for all available searches for non-standard Higgs bosons.  The Higgs decay widths, including the state-of-the art higher
order QCD corrections were calculated with {\tt sHDECAY}~\cite{Costa:2015llh} \footnote{The program {\tt sHDECAY} can be downloaded
  from the url: \url{http://www.itp.kit.edu/~maggie/sHDECAY}.}. {\tt sHDECAY} is based on the implementation of the models in
{\tt HDECAY}~\cite{Djouadi:1997yw,Butterworth:2010ym}. In our calculations all electroweak radiative corrections are turned off 
for consistency. A detailed description of the program can be found in appendix A of \cite{Costa:2015llh}. 

Up until run 2 of the LHC only LEP had constraints on resonances below 100 GeV after the discovery of the Higgs boson. 
New analyses using data from the run 2 of the LHC, at 13 TeV, constrain now the production and decay rates of heavy resonances 
into gauge bosons~\cite{Aaboud:2017rel,Aaboud:2017gsl,Aaboud:2018bun} and into SM-like Higgs pairs~\cite{Aaboud:2018knk,Aaboud:2018sfw}. 
These bounds are included in our analysis. The most stringent of the bounds is by far the one from~\cite{Aaboud:2017rel}. However, as we will
show, even if a substancial number of points are excluded by this analysis, the values of $\sin \alpha$ allowed by the very strong constraints
on the 125 GeV Higgs couplings to SM particles (modified by the common factor $\cos \alpha$ that affects all couplings in the same way) are
roughly in the same region.  

%

For the DM phenomenology, we consider the constraints from the cosmological DM relic abundance, collider searches, DM direct and indirect detections. The DM relic abundance for each model is calculated with the \texttt{MicrOMEGAs} code~\cite{Belanger:2013oya}, which is compared with the current experimental result $({\Omega}h^2)^{\rm obs}_{\rm DM} = 0.1186 \pm 0.002$ from the Planck Collaboration~\cite{Ade:2015xua}. Note that here we do not restrict the DM relic abundance to be exactly at the experimental value. Rather, we only require the model predicted value be equal to or smaller than the observed one. This way, we can consider both the dominant and subdominant DM cases simultaneously, for which we define the following DM fraction
\begin{eqnarray}
f_{A,X} = \frac{({\Omega} h^2)_{A,X}}{(\Omega h^2)^{\rm obs}_{\rm DM}}\,, 
\end{eqnarray} 
where $(\Omega h^2)_{A,X}$ denote the calculated DM relic abundance for either the pseudoscalar DM $A$ or the VDM $X$. 

The Higgs portal couplings can induce spin-independent DM-nucleon recoils for both scalar and vector DM models and the corresponding expressions have already been presented in Eqs.~(\ref{sigAN}) and (\ref{sigXN}). Currently, the LUX~\cite{Akerib:2016vxi}, PandaX-II~\cite{Cui:2017nnn} and XENON1T~\cite{Aprile:2017iyp,Aprile:2018dbl} experiments give the most stringent upper bounds for the DM nucleon scattering. In our work, we apply the latest XENON1T upper bounds~\cite{Aprile:2018dbl} for DM mass greater than 6~GeV, while for lighter DM particles, the combined limits from CRESST-II~\cite{Angloher:2015ewa} and CDMSlite~\cite{Agnese:2015nto} are used. Note that these experimental DM-nucleon scattering upper limits were derived by assuming that the DM candidate comprises all of DM abundance. Therefore, the proper quantity to be directly compared with experimental limits should be the effective DM-nucleon cross-section defined by $\sigma^{\rm eff}_{AN,XN} \equiv f_{A,X} \sigma_{AN, XN}$.

The DM indirect detection experiments can also impose strong constraints on the DM properties. In the models considered in the present work, the annihilation of DM into visible states may manifest itself in: the temperature anisotropies of CMB radiation, the $\gamma$-ray signals in the spheroidal dwarf galaxies or as the $e^\pm$ excesses in the Milky Way what can be probed and constrained by the observations of Planck~\cite{Ade:2015xua}, Fermi-LAT~\cite{Ackermann:2015zua} and AMS-02~\cite{Aguilar:2014fea,Accardo:2014lma}, respectively. According to Ref.~\cite{Elor:2015bho}, it is shown that for the DM mass range of interest, the Fermi-LAT upper bound on the DM annihilations from dwarfs is the most stringent. Note that both for the scalar and vector DM models, most of DM annihilations through the Higgs portal goes into $ZZ$, $W^+ W^-$, $b\bar{b}$ and light quark pairs. According to Ref.~\cite{Ackermann:2015zua}, all of these final states give nearly the same upper limits on the DM annihilation cross-sections. Thus, we use the Fermi-LAT bound from Ref.~\cite{Ackermann:2015zua} on $b\bar{b}$ when $m_{A,X} \geqslant m_b$, and that on light quarks for $m_{A,X}< m_b$. Also, similar to the DM direct detections, the comparison with the data requires the use of the effective DM annihilation cross-sections defined by $\sigma^{\rm eff}_{AA,XX} = f^2_{A,X} \sigma_{AA,XX}$, which are computed with the \texttt{MicrOMEGAs} code~\cite{Belanger:2013oya} automatically. 

Collider searches can provide information on DM particles through the SM-like Higgs $h_1$ invisible decay, with the corresponding decay width given in Eqs.~(\ref{widAA}) and (\ref{widXX}) for both DM models. The predicted Higgs invisible decay branching ratios should be compared with the LHC bound on this channel ${\rm Br}(h_1 \to {\rm inv}) = 0.24$~\cite{Patrignani:2016xqp}.

\section{Results}
\label{results}
In this section we compare the available parameter space for the two models after applying all the constraints described in section~\ref{numericalsim}. 
Again we note that the models have the same number of independent parameters. From the phenomenological point of view, the experimental measured 
quantities are the same, the second Higgs mass, the DM mass, the mixing angle $\alpha$ and the singlet VEV.
It is clear that the LHC cannot prove the existence of DM if it is not confirmed by direct detection experiments. It
is also true that the existence of a second neutral Higgs is predicted in most of the simplest extensions of the scalar sector. However, if a new scalar is discovered
while a hint for DM appears in the form of say, mono-X events, it may be possible to exclude some DM models if the events are in a region of
the parameter space already excluded. In the remainder of this section the colour code in the figures is the following:  red is for scalar DM and blue for vector DM, and in both cases relic density is not saturated, meaning that extra DM candidates are needed; on top of those points we present the points that are within 5$\sigma$ of the central value of the relic abundance value, in pink for the scalar case and in purple for the vector case. The colours are superimposed in the following order: red, blue, pink and then purple (so for instance a red dot may be hidden behind a blue dot).
\begin{figure}[h!]
	\centering
	\includegraphics[width=0.65\textwidth]{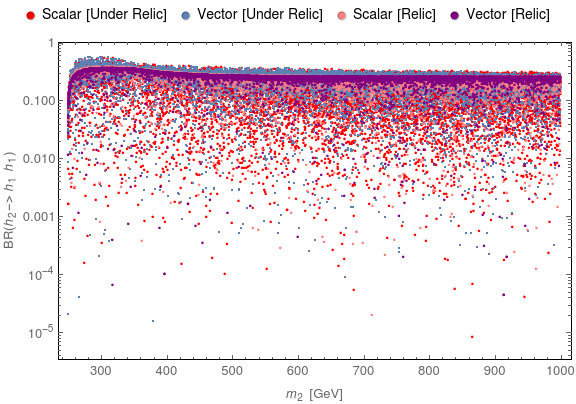}
	\vspace*{0.5cm}
	\caption{Branching ratio of $h_2 \to h_1 h_1$ as a function of $m_2$ for the scalar model and for the vector model (colour code in the legend).}
	\label{fig:BR_H_into_hh}
\end{figure}

We start with fig.~\ref{fig:BR_H_into_hh} where we present the branching ratio of $h_2 \to h_1 h_1$ as a function of $m_2$ for the scalar model and for the vector model.
Clearly, there is no significant difference between the two models. Values for the branching ratio reach a maximum of 70\% just after the channel opens and
then reduces to maximum values of about 40\%. However, if relic density is saturated the branching ratio is mostly below 40\% and again indistinguishable 
for the two models.  
 
%
\begin{figure}[h!]
	\begin{subfigure}{0.5\textwidth}
		\centering
		\includegraphics[width=1\linewidth]{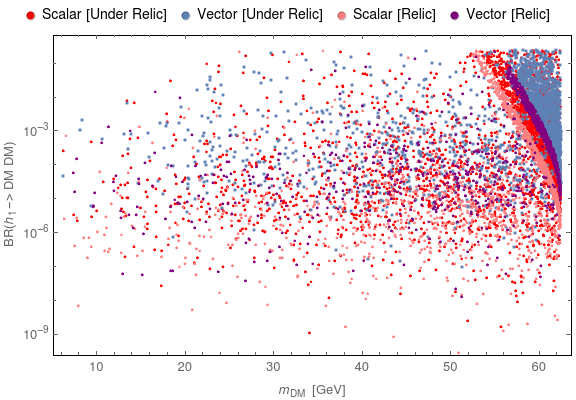}
		\caption{SM-like Higgs.}
	\end{subfigure}
	\begin{subfigure}{0.5\linewidth}
		\centering
		\includegraphics[width=1\textwidth]{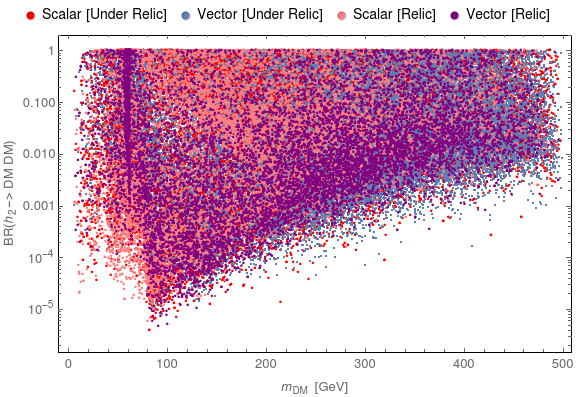}
		\caption{Second Higgs.}
	\end{subfigure}
	\vspace*{0.5cm}
	\caption{Branching ratio of the SM-like Higgs (a) and of the second Higgs (b) into DM particles as a function of the DM mass.}
	\label{fig:BR_h_into_xx}
\end{figure}
In Fig.~\ref{fig:BR_h_into_xx} we plot the branching ratio of the SM-like Higgs (a) and of the second Higgs (b) into DM particles as a function of the DM mass. 
Once more no significant deviations can be seen between the models and in this case there is no difference from the saturated to the non-saturated scenario.

\begin{figure}[h!]
	\begin{subfigure}{0.5\textwidth}
		\centering
		\includegraphics[width=1\linewidth]{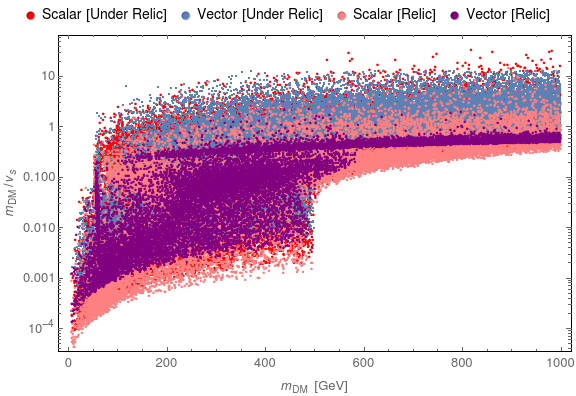}
	\end{subfigure}
	\begin{subfigure}{0.5\linewidth}
		\centering
		\includegraphics[width=1\textwidth]{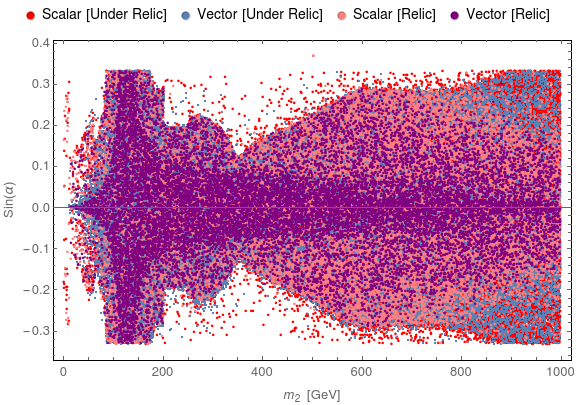}
	\end{subfigure}
	\vspace*{0.5cm}
	\caption{Left: $m_{DM}/\vs$ as a function of the DM mass; right: $\sin \alpha$ as a function of $m_2$.}
	\label{fig:sin_dark_m2}
\end{figure}

\begin{figure}[!h]
	\centering
	\includegraphics[width=0.55\textwidth]{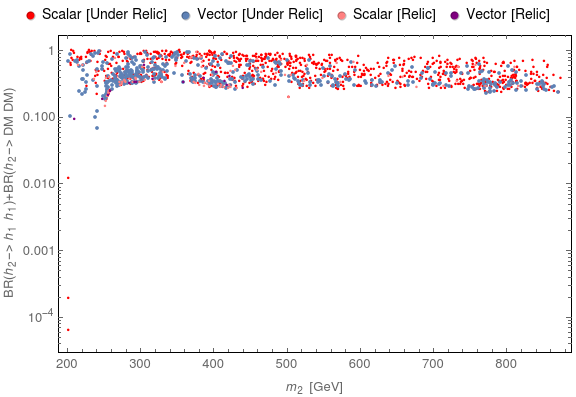}
	\vspace*{0.5cm}
	\caption{BR($h_2 \to h_1 h_1$)+ BR($h_2 \to \text{DM} \, \text{DM}$) vs. $m_2$ for points that survive the bounds coming from 
	heavy resonances and in particular $\sigma(pp(gg)\to h_2\to ZZ)$ with still large values $\sin \alpha$. Only points located outside of the pattern in the right panel of fig.~\ref{fig:sin_dark_m2} are shown.
	}
	\label{fig:surviving_points_branching}
\end{figure}

In the left panel of Fig.~\ref{fig:sin_dark_m2} we plot $m_{DM}/\vs$, a quantity that reduces to the gauge coupling constant in the VDM model:
\begin{equation}
\frac{m_{DM}}{v_S} =
\left\{
\begin{array}{lc}
\gx & \text{for VDM}\\
&\\
\frac{m_A}{\vs} & \text{for SDM}
\end{array} 
\right. 
\non
\end{equation} 
Roughly the same region is populated by both models. 
Note that points with suppressed $m_{DM}/\vs$ in the range between $10^{-4}$ and $10^{-2}$ for $m_{DM}\lsim 500\gev$ correspond
to $h_2$ resonances. In our scan $m_2$ varies between $1$ and $1000\gev$, therefore the resonances ($m_{DM}\sim m_2/2$) are distributed nearly uniformly 
for  $1\gev \lsim m_{DM}\lsim 500\gev$. For those points the requirement of proper DM abundance imply suppression of the coupling between DM and the resonance, so that $m_{DM}/\vs$ must be small.

In the right panel of Fig.~\ref{fig:sin_dark_m2} we show $\sin \alpha$ as a function of the second Higgs mass. The allowed band 
between about $-0.34$ and $0.34$ for $m_2$ above roughly $m_1/2$ is a hard ($m_2$-independent) bound on $\sin \alpha$ that comes from the combined signal strength measurements of the production and decay of the SM-like Higgs, $h_1$. This bound is weaker than in case of the real singlet model
with no DM candidate. The reason is that both BR($h_1 \to VV$) and BR($h_1 \to f\bar f$) might be reduced if  $\mdm< m_1/2$. 
The ``pattern'' of densely populated points is visible in the right panel of fig.~\ref{fig:sin_dark_m2}. Those points are originating from constraints imposed by searches for heavy scalar resonances, i.e. $h_2$ in our case. The pattern originates mainly from the search for $pp(gg)\to h_2\to ZZ$. Let us focus on this final state.
The $h_2$ production cross section is the same as for the SM multiplied by $\sin^2 \alpha$. Therefore, the shape of the pattern reflects mostly the behavior of the SM cross-section $\sigma(h_{SM})$ as a function of $m_{SM}$ that is here replaced by $m_2$. That is the reason why the exclusion is maximal close to the $t \bar t$ threshold, where the Higgs production cross section via gluon fusion has a local maximum.
Notice also the presence of less densely populated regions outside of the pattern with relatively large values of $\sin \alpha$. 
In order to understand its appearance, one should note that the total width of the second Higgs has an extra contribution $\Gamma(h_2 \to \text{DM} \, \text{DM})$. Therefore,
in contrast to what happens in the singlet extension with no dark matter candidate, here the BR($h_2 \to ZZ$) might be suppressed implying larger allowed values of $\sin\alpha$ located outside of the pattern.
To illustrate this point we plot in Fig.~\ref{fig:surviving_points_branching}  BR($h_2 \to \text{DM} \, \text{DM}$) + BR($h_2 \to h_1 h_1$) 
as a function of $m_2$ but only for points outside of the pattern. As expected, all those points correspond to large value of BR($h_2 \to \text{DM} \, \text{DM}$) + BR($h_2 \to h_1 h_1$) . 
The reason to have much fewer points outside of the pattern is that the decay $h_2 \to \text{DM} \, \text{DM}$ has to be allowed while the range 
of variation of $m_2$ and of the $m_{DM}$ is the same. That eliminates 3/4 of points in the considered region.

Finally, one can clearly see the result of the searches for $h_2 \to h_1 h_1$ close to 
the cross-section threshold and also the much harder bound for $m_2 < m_1/2$. Regarding the comparison of the two models we again see no difference
and the same can be said for the projection in the $(\sin \alpha, m_2)$ plane.

\begin{figure}[!h]
	\centering
	\includegraphics[width=0.65\textwidth]{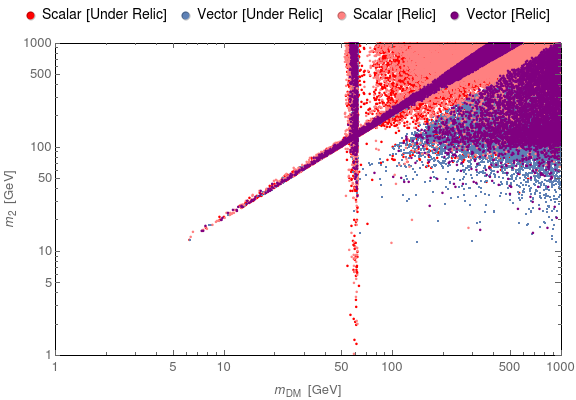}
	\vspace*{0.5cm}
	\caption{Second Higgs mass ($m_2$) as a function of the DM mass ($m_{DM}$.}
	\label{fig:coupling_constant}
\end{figure}

\begin{figure}[!h]
	\centering
	\includegraphics[width=0.65\textwidth]{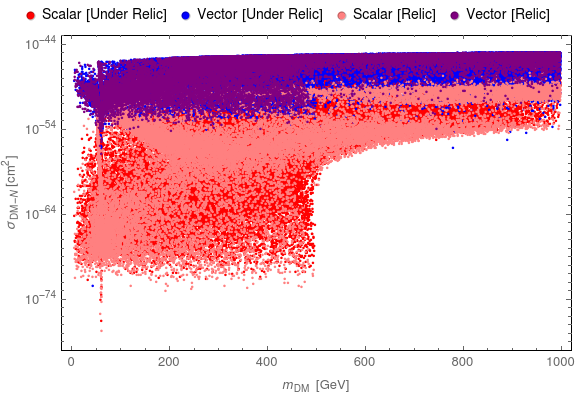}
	\vspace*{0.5cm}
	\caption{DM-nucleon cross-section as a function of the DM mass. Scalar DM-nucleon nucleon cross-section is computed at one-loop level. The latest results from Xenon1T are shown as the solid line that makes the upper edge of the plot.}
	\label{fig:XN_crosssection_oneloop}
\end{figure}

\begin{figure}[h!]
	\begin{subfigure}{0.5\textwidth}
		\centering
		\includegraphics[width=1\linewidth]{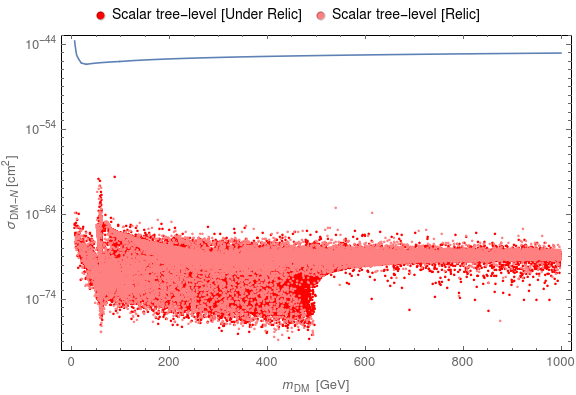}
	\end{subfigure}
	\begin{subfigure}{0.5\linewidth}
		\centering
		\includegraphics[width=1\textwidth]{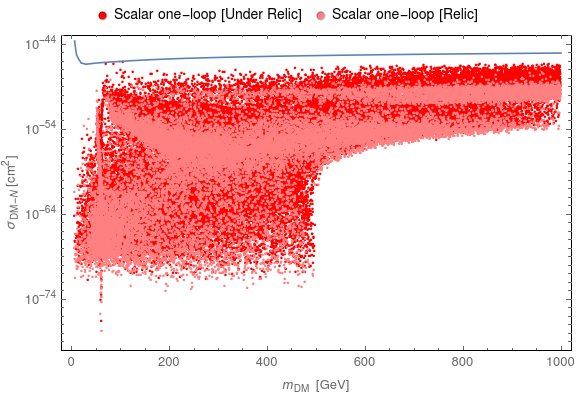}
	\end{subfigure}
	\vspace*{0.5cm}
	\caption{Scalar DM-nucleon cross-section as a function of the DM mass ($m_{DM}=m_A$) with the latest result from Xenon1T
	and relic abundance within $5\sigma$ of experimental value. 
	}
	\label{fig:sinalpha_m2_mdm}
\end{figure}

In Fig.~\ref{fig:coupling_constant} we show $m_2$ as a function of the DM mass. This is a projection of the parameter space where a clear difference between the two models can be seen. There are two bands where the models coexist, close to $m_{DM} \simeq m_1/2$ and to  
$m_2 \sim 2\cdot m_{DM}$. The explanation for the band structure could be easily guessed; these are the two resonances $h_1$ and $h_2$, respectively. In those regions, the kinematical enhancement by a resonance must be compensated by suppressed couplings that govern DM annihilation in the early Universe. This mechanism is nearly the same in both models. However, as seen from the figure there are two distinct regions above and below $m_2 = 2\cdot m_{DM}$ where only the scalar model survives. Hence, there are pairs of values $(m_2, m_X)$ that if hinted at the LHC will allow to exclude the vector model in favour of the scalar one. The reverse is not true as can be seen from the figure. 
The absence of VDM points in those regions is clarified in Fig.~\ref{fig:XN_crosssection_oneloop}, where a large suppression of the cross-section for scalar DM-nucleon scattering relative to the vector model one can be seen. In fact, a large portion of the parameter space of the VDM is excluded because they are above the Xenon1T bound. Therefore for a given $m_{DM}$ there exist $m_2$ large enough to be excluded by the Xenon1T bound. On the other hand,  for the SDM, even including one-loop corrections~\footnote{Hereafter, in this context, we are referring to the estimate of the upper bound for one-loop radiative corrections as given in (\ref{sigAN}).}, all points are below the Xenon1T line.  
In order to have a clear picture of what happens for the SDM we should compare the effect of one-loop versus tree-level result.  This is shown in    
Fig.~\ref{fig:sinalpha_m2_mdm} where in the left panel we show the result for the tree-level cross-section and in the right panel we show the one-loop result using equation~\ref{sigAN}. At tree-level the cross-section are more than orders of magnitude below the Xenon1T line. This is due to the nature of the scalar DM coupling to the Higgs bosons for which a detailed account is given in the appendix. The inclusion of the one-loop contributions for the SDM increases the maximum values of the cross-section by roughly ten orders of magnitude. Still only a few points are close to Xenon1T represented by the solid line (the upper edge) in the plots. Therefore, the SDM is still not affected by the direct constraints even with the one-loop corrections. Note that the points with maximally suppressed cross-section correspond to $h_2$ resonances scattered in the range 
$1\gev \lsim m_{DM} \lsim 500\gev$.

\begin{figure}[!h]
	\centering
	\includegraphics[width=0.65\textwidth]{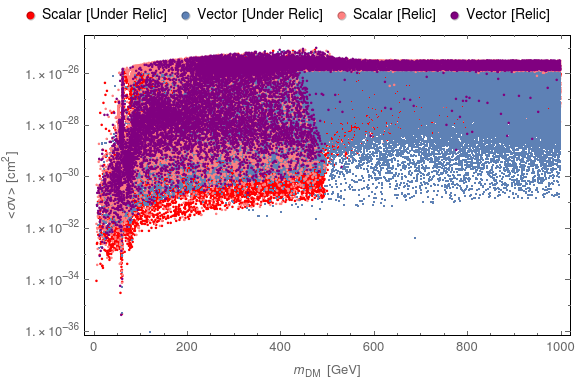}
	\vspace*{0.5cm}
	\caption{Thermal average DM annihilation cross-section (into the SM) times velocity (at zero temperature) versus DM mass.}
	\label{fig:sigmav_crosssection}
\end{figure}

Finally, we show in Fig.~\ref{fig:sigmav_crosssection} thermal average DM annihilation cross-section into the SM times velocity (at zero temperature) versus DM mass. Contrary
to the direct bound, the indirect bound affects both the SDM and the VDM. Although the density of points varies, the fact is that there are no major differences between the two models.
Furthermore  the allowed points for both models span a very large range of cross-sections and therefore will most probably not be excluded in the near future.

\section{Summary and conclusions}
\label{sec:conclusion}

In this paper we have compared scalar and vector dark matter models, both originating from the extension of the SM with an extra complex scalar $S$  and $\uone$ symmetry. In the first model, the global $\uone$ is softly broken by the term that generates the mass of the DM candidate which otherwise remains a massless Goldstone boson. In the second case, the $\uone$ is local and broken spontaneously therefore this massless mode contributes, within the Higgs mechanism, to the massive vector DM particle.
 
We have investigated the possibility to differentiate the models by measuring the energy distribution of $Z$ bosons at the ILC in the process $e^+e^- \to Z + \text{DM}$.
The final conclusion requires a dedicated experimental analysis which takes into account the full background and experimental details of the collider and the detector, a task which is far beyond the scope of this project. 
However, theoretical predictions show that there are regions in the $(m_{DM},m_2)$ space for which the total cross-section  predicted within the SDM is nearly 50\% larger than the one for the VDM, so that in those regions, future electron-positron colliders such as the ILC, CLIC or CEPC are likely to be a helpful tool in disentangling the two models. Unfortunately, the Standard Model $Z\nu_l\bar\nu_l$ final sates constitute background of considerable magnitude, making $e^+e^-$ collider search for DM very challenging.

We have shown that the direct detection is efficiently suppressed in the SDM model, $\sigma_{DM-N} \propto v^4_A$, as a consequence of $A$ being a pseudo-Goldstone boson.  The inclusion of one-loop corrections in the direct detection cross-section increases its maximum values by roughly ten orders of magnitude. Still, the bounds on direct detection do not affect the SDM. 

We have determined regions in the $(m_{DM},m_2)$ space 
that are excluded for the VDM while being allowed for the SDM. If future measurements point to those regions, the VDM will not be a viable option for DM. Those regions are excluded in the VDM since the DM-nucleon scattering in this case is not particularly suppressed and therefore consistency with Xenon1T eliminates a substantial part of the VDM parameter space. In the SDM the scattering is naturally very much suppressed, and the mechanism of the suppression has been explained in a more general context.


\section*{Acknowledgements}
We thank Pyungwon Ko for informing us about the paper~\cite{Gross:2017dan} by Gross, Lebedev and Toma.
This work is supported in part by the National Science Centre (Poland), research projects no~2014/15/B/ST2/00108, no~2017/25/B/ST2/00191 and a HARMONIA project under contract UMO-2015/18/M/ST2/00518 (2016-2019).

\appendix
\section{Goldston-boson--Higgs-boson coupling in a linear formalizm}
\label{app_A}
In order to gain a better understanding of the cancellation observed in sec.~\ref{DD_dd_sdm} we derive the coupling 
between two Goldstone bosons and a Higgs boson in a slight more general context. In this appendix we adopt the linear formalism.

Assume that the potential is composed by an invariant part, $V_\text{inv}$, and a softly breaking part $V_\text{soft}$, 
under certain symmetry transformation $G$
\beq
\phi_i \to \phi_i + \delta\phi_i = \phi_i + i \theta^a T^a_{ij}\phi_j,
\label{trans}
\eeq
where $T^a$ are the generators of the Lie algebra of the group $G$ and $\theta^a$ are the corresponding parameters.
So that
\beq
\delta V = \frac{\partial V}{\partial \phi_i} \delta \phi_i =
\frac{\partial V_\text{soft}}{\partial \phi_i} \delta \phi_i,
\label{var_con}
\eeq
with $V=V_\text{inv}+V_\text{soft}$. We assume $\phi_i$ are real fields. 
Explicitly one can write  
\beq
\frac{\partial V}{\partial \phi_i} \theta^a T^a_{ij}\phi_j =
 \frac{\partial V_\text{soft}}{\partial \phi_i} \theta^a T^a_{ij}\phi_j 
\label{inv_con}
\eeq
Differentiating twice with respect to $\phi_k$ and $\phi_l$ and evaluating the final expression at a minimum $\phi_n=\langle \phi_n\rangle = v_n$ of the full theory, i.e. for $V=V_\text{inv}+V_\text{soft}$, one obtains
\beq
V_{lki}\theta^a T^a_{ij}v_j + \left\{M^2_{ki} \theta^a T^a_{il} + (k \leftrightarrow l) \right\} = 
\left.\frac{\partial^3 V_\text{soft}}{\partial \phi_l \partial\phi_k \partial\phi_i}\right|_{\phi_n=v_n} \theta^a T^a_{ij}v_j + 
\left.\left\{\frac{\partial^2 V_\text{soft}}{\partial \phi_k \partial\phi_i}\right|_{\phi_n=v_n} \theta^a T^a_{il} + (k \leftrightarrow l) \right\},
\label{rel}
\eeq
where
\beq
V_{lki} \equiv \left. \frac{\partial^3 V}{\partial \phi_l \partial\phi_k \partial\phi_i}\right|_{\phi_n=v_n}\lsp\text{and}\lsp
M^2_{ki} \equiv \left.\frac{\partial^2 V}{\partial \phi_k \partial\phi_i}\right|_{\phi_n=v_n}.
\eeq
We shall specialise to the case of a complex singlet $S$ charged under a $U(1)$ symmetry
\beq
\phi =
\left( 
\begin{array}{c} \phi_1 \\ \vdots \\ \phi_{N-2} \\ s\equiv \frac{\Re S}{\sqrt{2}} \\ a\equiv \frac{\Im S}{\sqrt{2}} \end{array} 
\right)
\hsp
v = \langle\phi\rangle =
\left( 
\begin{array}{c} v_1 \\ \vdots \\ v_{N-2} \\ v_S \\ 0 \end{array} 
\right)
\hsp
M^2 =
\left( 
\begin{array}{cccc} 
M_{1,1}^2 & \cdots & M_{1, N-1}^2 & 0 \\
\vdots & \ddots & \vdots & \vdots \\
M_{N-1,1}^2 & \cdots & M^2_{N-1,N-1} & 0 \\
0 & \cdots & 0  & m^2_a
\end{array} 
\right)
\eeq
Note that the mass matrix $M^2$ is, in general, non-diagonal. Since we assume invariance under $S\to S^*$,  there is no mixing between $\Im S$ and other states in the mass matrix if $\langle a \rangle =0$. Since the $U(1)$ is softly broken the $a$ mass could be non-zero, i.e., a pseudo-Goldstone boson.

The $U(1)$ generator in this basis reads
\beq
T =
\left( 
\begin{array}{cccc} 
0 & \cdots & 0 & 0 \\
\vdots & \ddots & \vdots & \vdots \\
0 & \cdots & 0 & i \\
0 & \cdots & -i  & 0
\end{array} 
\right) \, .
\label{gen}
\eeq
In other words
\beq
T_{il}=i(\delta_{i,N-1}\delta_{l,N}-\delta_{i,N}\delta_{l,N-1}) \, ,
\eeq
so that 
\beq
M^2_{ki} T_{il}=i(M^2_{k, N-1}\delta_{l,N}-M^2_{k,N}\delta_{l,N-1}) \, ,
\hsp \text{and} \hsp
T_{ij}v_j= -i \delta_{i,N} v_S \, .
\eeq
Replacing the above in (\ref{rel}) and choosing the $V_{lNN}$ component one finds
\beq
V_{lNN}v_S =  M^2_{l,N-1} -\delta M^2_{l,N-1}  - (m_a^2  - \delta M^2_{N,N}) \delta_{l,N-1}
+\left.\frac{\partial^3 V_\text{soft}}{\partial \phi_l \partial\phi_N \partial\phi_N}\right|_{\phi_n=v_n} v_S,
\label{res2}
\eeq
where 
\beq
\delta M^2_{k,i}\equiv \left.\frac{\partial^2 V_\text{soft}}{\partial \phi_k \partial\phi_i}\right|_{\phi_n=v_n}.
\eeq
Note that if $V_\text{soft} \neq 0$ then $m^2_a$ receives contributions from the symmetric part of the potential as well~\footnote{Of course, those contributions vanish in the limit $V_\text{soft} \to 0$.} and therefore
\beq
m_a^2 \equiv \left.\frac{\partial^2 V}{\partial \phi_N \partial\phi_N}\right|_{\phi_n=v_n} \neq \left.\frac{\partial^2 V_\text{soft}}{\partial \phi_N \partial\phi_N}\right|_{\phi_n=v_n} \equiv \delta M^2_{N,N}
\eeq
In the symmetric limit of $V_\text{soft} \to 0$ one obtains $V_{lNN}v_S =  M^2_{l,N-1}$.
Note that the contribution $m_a^2  - \delta M^2_{N,N}$ might be written also in the following way
\beq
m_a^2  - \delta M^2_{N,N} = \left.\frac{\partial^2 V_\text{inv}}{\partial \phi_N \partial\phi_N}\right|_{\phi_n=v_n},
\label{g_mass}
\eeq
where $v_i$ in the vacuum of the full theory, i.e. for $V=V_\text{inv}+V_\text{soft}$.
  
The mass matrix $M^2$ could be diagonalized by an orthogonal rotation $R$ as follows 
\beq
M^2=R {\cal M}^2 R^T,
\eeq 
where ${\cal M}^2$ is the diagonal matrix. The mass eigenstes are $\varphi= R^T \phi$.
The rotation matrix is of the form
\beq
R=
\left( 
\begin{array}{cccc} 
R_{1,1} & \cdots & R_{1,N-1} & 0 \\
\vdots & \ddots & \vdots & \vdots \\
R_{N-1,1} & \cdots & R_{N-1,N-1} & 0 \\
0 & \cdots & 0 & 1
\end{array} 
\right) \, .
\eeq
The cubic coupling that is relevant for us could be written in terms of the mass eigenstates as follows
\beq
V = \cdots +  V_{lki} R_{ll'}\varphi_{l'} R_{kk'}\varphi_{k'} R_{ii'}\varphi_{i'} + \cdots
\label{pot3a}
\eeq
We are interested in the $V_{lNN}$ vertex and therefore we choose $k'=i'=N$. We also limit ourself to $l'\neq N$.
Since $R_{k,N}=\delta_{k,N}$
and $R_{i,N}=\delta_{i,N}$,  
\beq
V = \cdots +  V_{lNN} R_{ll'}\varphi_{l'}\varphi_{N} \varphi_{N} + \cdots
\label{pot3b}
\eeq
The term $M^2_{l,N-1}R_{ll'}$ from (\ref{res2}) together with (\ref{pot3b}) can be expressed by mass eigenvalues and mixing angles as 
$M^2_{l,N-1}R_{ll'} = m_{l'}^2 R_{N-1,l'}$. Then the coefficient of $\varphi_{l'} \varphi_{N} \varphi_{N}$ (with $l'\neq N$) reads
\beq
\frac{1}{v_S} 
\left\{m_{l'}^2 R_{N-1,l'} \; +   
\left.\left[
  \frac{\partial^3 V_\text{soft}}{\partial \phi_l \partial\phi_N \partial\phi_N} v_S R_{l,l'}
- \frac{\partial^2 V_\text{inv}}{\partial \phi_N \partial \phi_N} R_{N-1,l'}
- \frac{\partial^2 V_\text{soft}}{\partial \phi_l \partial\phi_{N-1}}R_{l,l'}\right]\right|_{\phi_n=v_n} 
\right\}
\label{ful_res}
\eeq
The above equation allows to calculate corrections to the $U(1)$-symmetric relation  $V_{l'NN} =  M^2_{l',N-1}/v_S$ for a given symmetry-breaking potential $V_\text{soft}$. For instance for $V_\text{soft}=\mu^2(S^2+S^{*2})$ the first term in the bracket is trivially zero while the remaining ones sum to zero
\beq
\left.\left[- \frac{\partial^2 V_\text{inv}}{\partial \phi_N \partial \phi_N} R_{N-1,l'}
- \frac{\partial^2 V_\text{soft}}{\partial \phi_l \partial\phi_{N-1}}R_{l,l'}\right]\right|_{\phi_n=v_n} 
= (4\mu^2-2\mu^2-2\mu^2)R_{N-1,l'}=0
\eeq
That way we have reproduced the result of (\ref{GGH}). It is also worth to consider a linear $U(1)$ breaking, 
by $M^3(S+S^*)/\sqrt{2}$. In this case, even though derivatives of $V_\text{soft}$ do not contribute 
to corrections to $V_{l'NN} =  M^2_{l',N-1}/v_S$, the derivative of $V_\text{inv}$, as it is evaluated at the 
minimum of the full theory, does contribute:
\beq
\left.\frac{\partial^2 V_\text{inv}}{\partial \phi_N \partial \phi_N}\right|_{\phi_n=v_n}  = -\frac{M^3}{v_S}
\eeq
Therefore we conclude that soft $U(1)$ breaking terms other than the quadratic ones may spoil the proportionality of the coupling to the Higgs mass squared  observed in (\ref{GGH}).
\section{Pseudo-Goldstone-boson--Higgs-boson Couplings in the non-linear formalism}
\label{app_B}
In this appendix  we rederive the above effective pseudo-Goldstone-Higgs couplings within the non-linear realization of the same Lagrangian. Here we write down the complex field $S$ in the following form:
\begin{eqnarray}
S = \frac{1}{\sqrt{2}} (v_s + s) e^{ia/v_s}\,,
\end{eqnarray}
so that the $U(1)$ symmetric part of the potential does not contain couplings involving the Goldstone boson $a$ any more. Since $a$ is odd under the $Z_2$ symmetry transformation $S \leftrightarrow S^*$, it can be an appropriate DM candidate. The only terms that $a$ appears in are the kinetic and the $U(1)$ softly-breaking terms. We will consider linear and quadratic breaking as follows:  
\begin{eqnarray}\label{La}
{\cal L}_a &=& \partial^\mu S^* \partial_\mu S - \frac{M^3}{\sqrt{2}} (S+S^*) -\mu^2 (S^2+S^{*2}) \nonumber\\
&=& \frac{(v_s +s)^2}{2 v_s^2}\partial^\mu a \partial_\mu a - M^3 (v_s + s)\cos\left(\frac{a}{v_s}\right) -\mu^2 (v_s+s)^2 \cos\left(\frac{2a}{v_s}\right)\,\\
&\supset & \frac{1}{2}\partial^\mu a \partial_\mu a + \frac{1}{2}\left(4\mu^2 +\frac{M^3}{v_s}\right)a^2 + \frac{s}{v_s}\partial^\mu a \partial_\mu a + \left(\frac{4\mu^2}{v_s} + \frac{M^3}{2 v_s^2}\right) s a^2\,,\nonumber
\end{eqnarray}
from which we can easily read off the pseudoscalar DM mass squared as $m_a^2 = -4\mu^2 - M^3/v_s$, which is the same as that obtained within the linear realization of the $U(1)$ symmetry. 

Now we are going to show that the pseudo-Goldstone-Higgs vertex agrees with the result obtained  in the appendix \ref{app_A}. We focus on the following vertex involving partial derivatives of $a$
\begin{eqnarray}\label{deria}
&& \frac{1}{v_s} s \partial^\mu a \partial_\mu a = -\frac{1}{v_s} (\partial^\mu s \partial_\mu a)a - \frac{1}{v_s} s a \square a  \nonumber\\
&=& -\frac{1}{2v_s} \partial^\mu s \partial_\mu (a^2) + \frac{m_a^2}{v_s} s a^2 = \frac{1}{2v_s} (\square s) a^2 + \frac{m_a^2}{v_s} s a^2 \\
&=& \frac{1}{2v_s} (\sin\alpha \square h_1 +\cos\alpha \square h_2 ) a^2 + \frac{m_a^2}{v_s} s a^2 = -\frac{1}{2v_s}(\sin\alpha m_1^2 h_1 +\cos\alpha m_2^2 h_2 ) a^2 + \frac{m_a^2}{v_s} s a^2\,, \nonumber
\end{eqnarray}
where we have repeatedly used the integration by parts and exploited free equations of motion for $a$ and $h_{1,2}$, {\it i.e.} $\square a = -m_a^2 a$ and $\square h_i = -m_i^2 h_i$. By putting the final expression of Eq.~(\ref{deria}) into Eq.~(\ref{La}), we obtain
\begin{eqnarray}
{\cal L}_a &\supset& \frac{1}{2} (\partial^\mu a \partial_\mu a -m_a^2 a^2) - \frac{1}{2v_s} (\sin\alpha m_1^2 h_1 + \cos\alpha m_2^2 h_2) a^2 + \frac{1}{v_s} (4\mu^2 + \frac{M^3}{2v_s} + m_a^2) s a^2 \nonumber\\
&=& \frac{1}{2} (\partial^\mu a \partial_\mu a -m_a^2 a^2 ) - \frac{1}{2v_s} (\sin\alpha m_1^2 h_1 + \cos\alpha m_2^2 h_2) a^2 - \frac{M^3}{2v_s^2} s a^2\\
&=& \frac{1}{2} (\partial^\mu a \partial_\mu a -m_a^2 a^2) - \frac{1}{2v_s} (\sin\alpha m_1^2 h_1 + \cos\alpha m_2^2 h_2) a^2 - \frac{M^3}{2v_s^2} (\sin\alpha h_1 + \cos\alpha h_2) a^2\,. \nonumber
\end{eqnarray}
So indeed, the coupling is the same as obtained in the appendix \ref{app_A} and in sec.~\ref{DD_dd_sdm}. 


\bibliography{bib_vec_scal_DM}{}
\bibliographystyle{JHEP}

\end{document}